\documentclass[12pt]{article}

\usepackage{latexsym}
\usepackage{amssymb,amsfonts,amsmath}
\usepackage{graphicx} 
\usepackage{indentfirst}
\usepackage{bbm}
\usepackage{amssymb}
\usepackage{verbatim}
\usepackage{amsmath, amsthm,amssymb}
\usepackage{mathrsfs}
\usepackage{hyperref}
\usepackage{amsfonts}
\usepackage{dsfont}
\usepackage{cite}
\usepackage{xcolor}
\usepackage{slashed}
\renewcommand{\thefootnote}{\fnsymbol{footnote}}

\newcommand{\bsubeq}{\begin{subequations}}
\newcommand{\esubeq}{\end{subequations}}

\usepackage{amsmath,mathrsfs,graphicx,epstopdf,placeins,float}
\usepackage{booktabs}
\newcommand{\be}{\begin{equation}}
\newcommand{\ee}{\end{equation}}
\newcommand{\ba}{\begin{eqnarray}}
\newcommand{\ea}{\end{eqnarray}}
\newcommand{\p}{\prime}

\begin{document}

\begin{titlepage}
\begin{flushright}
Febuary, 2024
\end{flushright}
\vspace{5mm}

\begin{center}
{\Large \bf 
Charged massless scalar fields in a charged $C$-metric black hole: 
Exact solutions, Hawking radiation, and scattering of scalar waves
}
\end{center}

\begin{center}

{\bf
Ming Chen,
Gabriele Tartaglino-Mazzucchelli,
Yao-Zhong Zhang
} \\
\vspace{5mm}

\footnotesize{
{\it 
School of Mathematics and Physics, University of Queensland,
\\
 St Lucia, Brisbane, Queensland 4072, Australia}
}
\vspace{2mm}
~\\
\texttt{m.chen3@uq.net.au,g.tartaglino-mazzucchelli@uq.edu.au,yzz@maths.uq.edu.au
}\\
\vspace{2mm}

\end{center}

\begin{abstract}
\baselineskip=14pt

We study Hawking radiation and wave scattering of charged scalar fields in a charged $C$-metric black hole background. The conformally invariant wave equation for charged scalar fields can be separated into radial and angular parts, each with five singularities. We first show that the radial and angular equations can be respectively transformed into the general Heun equation, and then we explore exact solutions of the radial Heun equation in terms of the local Heun functions and connection coefficients. Exact behaviors of the asymptotic wave functions are determined without approximations. We further apply the exact results to derive Hawking radiation, quasinormal modes and superradiance. 
\end{abstract}
\vspace{5mm}

\vfill
\end{titlepage}

%%%%%%%%%%%%

\newpage
\renewcommand{\thefootnote}{\arabic{footnote}}
\setcounter{footnote}{0}

\tableofcontents{}
\vspace{1cm}
\bigskip\hrule

\allowdisplaybreaks
%%%%%%%%%%%%%%%%%%%%%%%%%%%%%%%%%
\section{Introduction}

The existence of black holes (BHs) is one of the most striking predictions in Einstein's theory of general relativity (GR). BHs are exact solutions of the Einstein field equations, characterized by elementary macroscopic quantities such as mass, angular momentum and electric charge. Their physics has been an immensely fruitful research ground, playing a vital role in examining gravitational fields under extreme conditions. Research on BHs has recently experienced substantial advancements, propelled by detecting gravitational waves from the neutron stars and black hole collisions \cite{ligo,raea}. Additionally, there have been groundbreaking measurements 
of the central BHs in the M87-galaxies \cite{m87}. 
These developments mark a significant leap forward in understanding the dynamics and properties of BHs in our cosmos. 

BHs were predicted to enjoy remarkable phenomena, such as Hawking radiation, when quantum effects come into play \cite{hawk4}. Under specific conditions, they can amplify incident waves at the cost of the black hole spins \cite{rprm} or charges \cite{jdb,kdun} through, e.g., the Penrose process and superradiance \cite{rbvcpp}.

Perhaps more significantly, the detected gravitational waves from a pair of merging BHs \cite{ligo} indicate that the interaction of two BHs can be conditionally divided into four stages: the Newtonian stage, the merger of two BHs into a single one, the ringdown phase, and the formation of a single final BH with the consequent ringing \cite{ligo,raea}.
The ringdown phase or ringdown waveform is originated from the distorted final resultant BH and comprises the superposition of quasinormal modes (QNMs).
Each QNM possesses a complex frequency, where the real part corresponds to the oscillation frequency and the imaginary part is the inverse of the damping time. This damping time is uniquely determined by the mass and angular momentum of the BH. 
As the frequencies and damping time of QNMs are directly related to the ``No Hair'' theorem of BHs, a precise identification of QNMs serves as a conclusive indicator for BHs and provides a crucial test for GR in the context of strong gravitational fields \cite{ebvccmw}.

Furthermore, it is well known that small perturbations of a BH background take the form of damped oscillations, which also lead to QNMs. 
By causality, QNMs can be calculated when the perturbation is purely ingoing at the exterior event horizon and is purely outgoing 
at spatial infinity \cite{ajo,kdk,bss,ebvcas}.
Various types of fields have been used as the test fields to induce perturbations in different BH backgrounds,
based on solutions of the Teukolsky master equations 
\cite{teu,teu2}.

In most cases, the master wave equations can be separated into a set of ordinary differential equations (ODEs). 
Various methods have been employed to solve these separated ODEs, both approximately and analytically
\cite{dvgn,yhhz,hskd,hsvc2,hsvc4,sqw,masadd}  - at least for subclasses of gravitational backgrounds that allow for that.
In recent years, more researchers have been interested in solving the ODEs analytically in terms of the Heun functions by transforming the ODEs into Heun's equations \cite{heun,smhset,hseh,dbhs,mhort,bgadd2}.
In \cite{hsvc4,hsvc3,hmsn}, the exact formulation for wave scatterings of BHs and Kerr-AdS$_5$ type spacetimes was presented based on the general Heun equation (GHE) and their exact solutions. 
In \cite{hskd,snhm,yh}, QNMs of the Kerr-dS BHs were studied in terms of the Heun functions.

In this paper, we develop and employ the analytic approach to study solutions and applications of the master equation in a charged $C$-metric background \cite{ccmetric}.
The $C$-metric serves as a mathematical model for describing a pair of BHs that are causally separated and moving apart from each other with opposite accelerations. This metric is an extension of the Schwarzschild solution and introduces an extra parameter associated with the acceleration of the BH, in addition to its mass parameter \cite{jb,kh}.
In the charged version of the $C$-metric, an electric charge parameter is included to account for the influence of an electromagnetic field. The $C$-metric represents a class of boost-symmetric BH geometry and can be interpreted as a BH that has been accelerated under its interaction with a local cosmological medium. 
The $C$-metric has also been studied in the context of supergravity, see, e.g., \cite{Klemm:2013eca,Nozawa:2022upa,Nozawa:2023aep}.

In \cite{cccmetric,ccccmetric}, the conformally invariant Klein-Gordon (KG) wave equation for a massless neutral scalar field in the charged $C$-metric BH was derived. Then \cite{ccmetric} further generalizes this coupling to the charged scalar field case. The master equations in both cases can all be separated into radial and angular parts (see Appendix A of \cite{ccmetric} as well as \cite{bgadd}). Using the asymptotic behaviors of the solutions of the separated ODEs and the Mathematica package QNMSpectral developed in \cite{ajo}, QNMs and superradiance of the $C$-metric BH were investigated in \cite{ccmetric}. 
These results indicated that the $C$-metric remains stable under the perturbation of neutral or charged scalar fields. This stability is reflected in the frequencies of QNM, which exhibit a monotonic decay over time.

In this work, we study Hawking radiation and wave scattering of a charged massless scalar field in the charged $C$-metric BH background. The corresponding conformally invariant KG wave equation separates into two ODEs with five singularities. We show that the radial and angular ODEs can be respectively transformed into the GHE. Exact solutions of the radial Heun equation are obtained in terms of the local Heun functions and connection coefficients. This enables us to determine the exact behaviours of the asymptotic wave functions. We apply the exact results to analyse Hawking radiation, QNMs and superradiance. 

This paper is organized as follows.
In Sec.\,\ref{ghe}, we derive the Heun equations describing the radial and angular parts of the KG wave equation in the $C$-metric background. In Sec.\,\ref{rghe}, we determine the asymptotic behaviors of the radial wave function and compute the Hawking radiation.
In Sec.\,\ref{bound}, we derive the exact coefficients of the asymptotic wave functions in the tortoise coordinates and apply them to discuss the boundary conditions for QNMs and superradiance. In Sec.\,\ref{gheeeb}, we present exact solutions of the Heun equations via connection coefficients and obtain an analytic formulation of QNMs and superradiance. We present the conclusions and discussions of the paper in Sec.\,\ref{sdos}.
In Appendix \ref{heun} we present more detail regarding the derivation of the radial Heun equation of Sec.\,\ref{ghe}.
Note that, throughout the paper, we set the gravitational constant and the speed of light equal to 1.

\section{Heun's equations}
\label{ghe}

The line element of the charged $C$-metric in the Boyer-Lindquist coordinates reads \cite{ccmetric,cccmetric,ccccmetric}
\begin{subequations}
\begin{eqnarray}\label{gm18}
ds^2&=&\frac{1}{\Lambda^2}\bigg(-f(r)dt^2+f^{-1}(r)dr^2+P(\theta)^{-1}r^2d\theta^2  
+P(\theta)r^2 \sin^2 \theta d\varphi^2 \bigg), 
\end{eqnarray}
where $\Lambda=1-\tilde{\alpha} r \cos \theta$ works as a conformal factor, $\tilde{\alpha}$ is the acceleration parameter which is positive real in the de Sitter spacetime, and the functions $f(r)$ and $P(\theta)$ are given by
\begin{eqnarray}
f(r)&=&\bigg(1-\frac{2M}{r}+\frac{Q^2}{r^2}\bigg)(1-\tilde{\alpha}^2 r^2),\\
P(\theta)&=&1-2\tilde{\alpha} M \cos \theta + \tilde{\alpha}^2 Q^2 \cos^2 \theta,
\end{eqnarray}
\end{subequations}
with $\theta \in (0,\pi)$ and $M$ being the mass of the $C$-metric BH. $Q$ is the electric charge, and the electromagnetic potential associated with the charged BH source is $A_\mu=(- Q/r, 0, 0, 0)$.

The conformally invariant wave equation for a massless charged scalar field $\phi$ in the $C$-metric background is given by \cite{ccmetric}
\begin{eqnarray}\label{rnc2}
0&=& \frac{\partial}{\partial r}\bigg(r^2 f(r) \frac{\partial \tilde{\phi}}{\partial r} \bigg)  -  \frac{r^4}{r^2 f(r)}\frac{\partial^2 \tilde{\phi}}{\partial t^2} - \frac{2iqQr}{r^2 f(r)}\frac{\partial \tilde{\phi}}{\partial t} + \frac{q^2Q^2}{r^2 f(r)}\tilde{\phi}  \nonumber \\
&&  + \frac{1}{6}\bigg( r^2f^{\p\p}(r) + 4rf^{\p}(r)+2f(r)  \bigg)\tilde{\phi}  + \frac{1}{\sin \theta}\frac{\partial}{\partial \theta}\bigg(  P(\theta) \sin \theta \frac{\partial}{\partial \theta} \tilde{\phi} \bigg)  \nonumber \\
&&  +  \frac{1}{P(\theta) \sin^2 \theta} \frac{\partial^2 \tilde{\phi}}{\partial \varphi^2} + \frac{1}{6} \bigg( P^{\p\p}(\theta) + 3\cot \theta P^{\p}(\theta) - 2P(\theta)  \bigg) \tilde{\phi}, 
\end{eqnarray}
where $\tilde{\phi}=\Lambda^{-1}\phi$ and $q$ is the charge of the scalar field. 

The conformally invariant wave equation (\ref{rnc2}) is separable \cite{dkfn}. Indeed, we consider the Ansatz 
\begin{equation}\label{rnc1add}
\tilde{\phi}(r,\theta,\varphi,t)=e^{-i \omega t} R(r) e^{i m \varphi}\,\Theta(\theta), 
\end{equation}
where $\omega$ is the frequency (energy) of the scalar field or wave under BH perturbations, and $m\in\mathbb{Z}$ is the azimuthal 
number. Moreover, $R(r)$ and $\Theta(\theta)$ are functions of the radial and angular variables $r$ and $\theta$, respectively 
\cite{synutf,azhi}.
Then, substituting (\ref{rnc1add}) into (\ref{rnc2}), we find that the wave equation is separated into the radial and angular ODEs,
\begin{subequations}
\begin{eqnarray}\label{rnc5}
R(r)^{\p\p}   +    \frac{\Delta^{\p}(r)}{\Delta(r)} R(r)^{\p}   +    \frac{1}{\Delta(r)}\bigg[\frac{\Omega(r)}{\Delta(r)}   +    \frac{1}{6}\Delta^{\p\p}(r)   +    \lambda \bigg] R(r)= 0,
\end{eqnarray}
\begin{eqnarray}\label{rnc4}
&&\Theta^{\p\p}(\theta)   +    \left(\frac{P^{\p}(\theta)}{P(\theta)}   +    \cot \theta \right)\Theta^{\p}(\theta)   +    \frac{1}{P(\theta)} \bigg[\frac{-m^2}{P(\theta)\sin^2 \theta}
 + 2 \tilde{\alpha} M \cos \theta  \nonumber \\
&&~~~~~~~~~~~~~~~~~~~~~~~~~~~~~~~~~~~ +    2\tilde{\alpha}^2 Q^2\cos^2 \theta   +    \frac{\tilde{\alpha}^2Q^2 - 1}{3}  +    \lambda \bigg] \Theta(\theta) = 0, 
\end{eqnarray}
\end{subequations}
where the prime denotes the derivative with respect to the argument,  $\lambda$ is the separation constant and     \begin{subequations}\begin{eqnarray}\label{del}
  \Delta(r)&=&r^2f(r)=-\tilde{\alpha}^2 (r^2-2Mr+Q^2)\bigg(r^2-\frac{1}{\tilde{\alpha}^2}\bigg),~~~
\\
\label{omee}
  \Omega(r)&=&\omega^2r^4 - 2qQ\omega r + q^2Q^2.  
\end{eqnarray} 
\end{subequations}

Before proceeding further, it is worth mentioning that when waves propagate near or cross the horizon of a BH, the extreme gravitational time dilation leads to a substantial change in the observed frequencies $\omega$. Typically, this effect leads to a discrete set of complex frequencies representing exponential decay of the waves if the imaginary part is negative, and dynamical instability of the waves if the imaginary part is positive \cite{ajo}. It is thus reasonable to assume that $\omega$ is complex, i.e. $\omega=\omega_{_R}+i \omega_{_I}$, where $\omega_{_R}=\textrm{Re}[\omega]$, $\omega_{_I}=\textrm{Im}[\omega]$, introduced above, are respectively the real and imaginary parts.
In the following we will use $J(r)$ to denote the square root of $\Omega(r)$, 
\begin{equation}J(r)= \sqrt{\Omega(r)},
\end{equation}
which can be expressed as
\begin{eqnarray}\label{app3}
J(r)= \bigg[\sqrt{\frac{\bar{r}+a}{2}}+i\;\textrm{sgn}(b) \sqrt{\frac{\bar{r}-a}{2}} \bigg], 
\end{eqnarray}
with $\bar{r}=\sqrt{a^2+b^2}$ being  the modulus, with $a=(\omega_{_R}^2-\omega_{_I}^2)r^4+ q^2Q^2-2qQ\omega_{_R} r$, and $b=2(\omega_{_R}\omega_{_I}r^4- qQ\omega_{_I} r)$.

The separated ODEs  \eqref{rnc5} and \eqref{rnc4} have five regular singularities. We now show that both equations can be transformed into the general Heun differential equation via the change of variables.

We start with the radial equation. Setting
\begin{eqnarray}\label{rerad1}
R(r)=\Delta^{-\frac{1}{2}}(r) \Psi(r),
\end{eqnarray}
we obtain from \eqref{rnc5} the ODE for $\Psi(r)$,
\begin{eqnarray}\label{rnc6}
\Psi^{\p\p}(r)  +   \bigg[ \frac{1}{4} \big(\frac{\Delta^{\p}(r)}{\Delta(r)}\big)^2    -  \frac{1}{3} \frac{\Delta^{\p\p}(r)}{\Delta(r)}  + \frac{\Omega(r)}{\Delta^2(r)}  +  \frac{\lambda}{\Delta(r)} \bigg]   \Psi(r) =  0
.~~~
\end{eqnarray}
This equation has four finite regular singularities as well as one infinite regular singularity at $r=\infty$. The finite singularities are determined by 
\begin{eqnarray}\label{rnc7}
\Delta(r)=-\tilde{\alpha}^2(r-r_{+})(r-r_{-})(r-r^{\p}_{+})(r-r^{\p}_{-})= 0,
\end{eqnarray}
where $r_{\pm}=M\pm \sqrt{M^2-Q^2}$ and $r^{\p}_{\pm}=\pm \frac{1}{\tilde{\alpha}}$. 
We set the ordering of the four roots as $r^{\p}_{-}< 0\leq r_{-} < r_{+} < r^{\p}_{+}$.
Then $r=r_\pm, r_+^{\p}$ are event, Cauchy and acceleration horizons, respectively, while $r=r_-^{\p}$ is not physical. We are interested in the static region $r_{+}< r < r^{\p}_{+}$, in which $\Delta(r)$ is positive and the metric has fixed signature \cite{cccmetric,hseh}. This ordering requires that $r_{+} \leq \frac{1}{\tilde{\alpha}}$, i.e., $\tilde{\alpha} (M+ \sqrt{M^2-Q^2})\leq 1$. When the equality holds, the BH becomes extremal. This case is also known as the Nariai limit. Another extremal limit occurs when $M=Q$. In this case, the event and Cauchy horizons coincide. Note that
in the static region $P(\theta)>0$ for all $\theta \in (0,\pi)$ \cite{bbacc,ccmetric}.

We make the following variable change 
\begin{eqnarray}\label{rnc8}
z :=\frac{r^{\p}_{+}-r_{-}}{r^{\p}_{+}-r_{+}} \frac{r-r_{+}}{r-r_{-}}. 
\end{eqnarray}
Under this transformation, 
\begin{subequations}    
\begin{eqnarray}\label{rnc8add}
& \;\;\;\;\;\;\;\;\;\;\;\;\;\;\;\;\;\;\;\; r=r_{+} \rightarrow  z=0,\;\;\quad r=r^{\p}_{+} \rightarrow  z=1, & \\
& \;\;\;\;\;\;\;\;\;\;\;\;\;\;\;\;\;\;\;\; r=r^{\p}_{-} \rightarrow  z=z_r, \;\;\quad  r=r_{-}\rightarrow  z=\infty, & 
\end{eqnarray}
\end{subequations}
where $z_r =z_{\infty}\frac{r^{\p}_{-}-r_{+}}{r^{\p}_{-}-r_{-}}$ with  $z_{\infty}=\frac{r^{\p}_{+}-r_{-}}{r^{\p}_{+}-r_{+}}$. 
Thus this transformation maps the region of interest, $r_{+}< r < r^{\p}_{+}$, to the one $0< z <1$. In terms of the new variable $z$, the radial equation \eqref{rnc6} becomes
\begin{eqnarray}\label{rnc14}
&&\Psi^{\prime\prime}(z) 
+
\frac{2}{z-z_{\infty}}\Psi^{\prime}(z)
\nonumber\\
&&~~~+\frac{(r_{+}-r_{-})^2z^2_{\infty}}{(z-z_{\infty})^4} \bigg[ \frac{1}{4} \bigg(\frac{\Delta^{\p}(r)}{\Delta(r)}\bigg)^2 
 - \frac{1}{3} \frac{\Delta^{\prime\prime}(r)}{\Delta(r)} +\frac{\Omega(r)}{\Delta^2(r)} + \frac{\lambda}{\Delta(r)} \bigg] \Psi(z)= 0. ~~~
\end{eqnarray}
Setting 
\begin{eqnarray}\label{rnc22}
\Psi(z)=z^{C_1}(z-1)^{C_2}(z-z_r)^{C_3}(z-z_{\infty})^{-1} \mathcal{Y}(z),
\end{eqnarray}
after tedious computations (see Appendix \ref{heun} for details), we arrive at the following radial ODE
\begin{eqnarray}\label{rnc36}
\mathcal{Y}^{\prime\prime}(z)  +   \bigg(   \frac{\gamma}{z} +   \frac{\delta}{z-1}  +  \frac{\epsilon}{z-z_r}   \bigg) \mathcal{Y}^{\prime}(z)
  +   \frac{\alpha\beta z-{q_r}}{z(z  -  1)(z  -  z_r)}  \mathcal{Y}(z)  = 0,
~~~
\end{eqnarray}
where \footnote{The equations in \eqref{appref}, \eqref{apprefadd} and \eqref{apppref} have double signs. These signs are all correlated, with only two branches of sign choices (for example, (+,-) and (-,+) in \eqref{apprefadd}). The choice of signs in either of these branches does not affect the outcome of asymptotic radial wave functions (see e.g. \eqref{rnc73} and \eqref{rnc74}).}
\bsubeq\label{apprefbs13}
\begin{eqnarray}\label{appref}
&&\gamma=2C_1, \;\;\quad \delta=2C_2,  \;\; \quad \epsilon=2C_3, \;\; \quad
\alpha=1 \pm \sum_{j=1}^4\tilde{B}_j, \\ \;\;
&&\beta= 1 \pm \sum_{j=1}^3\tilde{B}_j \mp\tilde{B}_4, \;\;\quad     
\label{apprefadd}
{q_r}=2C_1C_3+2C_1C_2 z_r-A z_r,   
\end{eqnarray}
\esubeq
with 
\bsubeq\label{apprefbs13add}
\begin{eqnarray}\label{apppref}
C_j=\frac{1}{2} \pm \tilde{B}_j, \;\;\quad  \tilde{B}_j :=  i  \frac{J(r_{j})}{\Delta^{\prime}(r_{j})}, 
\end{eqnarray}
\begin{eqnarray}
\label{rnc55}
A&=&\frac{1}{3}\bigg[ \frac{r^{\prime}_{+}-r_{+}}{r^{\prime}_{+}-r_{-}} +\frac{1}{2}\frac{r_{-}-r_{+}}{r_{-}-r^{\prime}_{+}}-\frac{1}{2}\frac{(r_{-}-r_{+})(r^{\prime}_{+}-r_{+})}{(r^{\prime}_{+}-r_{-})(r^{\prime}_{-}-r_{+})} \bigg] \nonumber \\
&&+\frac{1}{z^2_r}\bigg(2E+D+\frac{2E}{z_r}\bigg) -\frac{\lambda}{\alpha^2}\frac{1}{(r_{-} - r^{\prime}_{+})(r_{-}-r^{\prime}_{-})}\frac{z_{\infty}}{z_r}, 
\\
E&=&\frac{J^2(r_{+})}{[\Delta^{\prime}(r_{-})]^2} z^4_{\infty}, \\
D&=& \frac{-4\omega^2r_{-}r^3_{+} + 2qQ\omega (r_{-}+3r_{+}) - 4q^2Q^2}{[\Delta^{\prime}(r_{-})]^2}z^3_{\infty}. 
\end{eqnarray}
\esubeq
Here and throughout, we  use the identification 
\begin{equation} r_1, r_2, r_3, r_4 \Longleftrightarrow
r_{+},r^{\prime}_{+},r^{\prime}_{-},r_{-},
\end{equation}
respectively.
It can be easily checked that  $\gamma+\delta+\epsilon=\alpha+\beta+1$. So the radial equation (\ref{rnc36}) is the GHE
with accessory parameter ${q_r}$. 

We now consider the angular part. Making the variable change $x=\cos \theta$ and setting
\begin{eqnarray}\label{reang1}
\Theta(x)= [(x^2-1)P(x)]^{\frac{1}{2}}\mathcal{X}(x),
\end{eqnarray}
where $P(x)=1-2\tilde{\alpha} Mx+\tilde{\alpha}^2 Q^2 x^2$, then the angular ODE \eqref{rnc4} becomes
\begin{eqnarray}\label{rnc37}
\mathcal{X}^{\p\p}(x)   +   \bigg[ \frac{1}{4} \bigg(\frac{\mathcal{P}^{\p}(x)}{\mathcal{P}(x)}\bigg)^2   -   \frac{1}{3} \frac{\mathcal{P}^{\p\p}(x)}{\mathcal{P}(x)}   +  \frac{(im)^2}{\mathcal{P}^2(x)}   +   \frac{\lambda}{\mathcal{P}(x)} \bigg] \mathcal{X}(x) =  0.
\end{eqnarray}
Here 
\bsubeq
\begin{eqnarray}
 \mathcal{P}(x)&=&(x^2-1)P(x)
 \\
 &\equiv&\tilde{\alpha}^2Q^2\,(x-x_{+})(x-x_{-})(x-x^{\p}_{+})(x-x^{\p}_{-}),\\  
 x_{+}&=&-1,\quad x_{-}=\frac{1}{\tilde{\alpha} Q^2} (M+\sqrt{M^2-Q^2}), \\
 x^{\p}_{+}&=&1,\quad
  x^{\p}_{-}=\frac{1}{\tilde{\alpha} Q^2} (M-\sqrt{M^2-Q^2}).
\end{eqnarray}
\esubeq

Let $ w^s_{\infty}=\frac{x^{\p}_{+}-x_{-}}{x^{\p}_{+}-x_{+}}$, $w_s=w^s_{\infty}\,\frac{x^{\p}_{-}-x_{+}}{x^{\p}_{-}-x_{-}}$. Similarly to the radial case, making the variable change
\begin{eqnarray}\label{rnc39}
w :=\frac{x^{\p}_{+} - x_{-}}{x^{\p}_{+} - x_{+}} \frac{x - x_{+}}{x - x_{-}},
\end{eqnarray}
and setting 
\begin{eqnarray}\label{angrnc22}
\mathcal{X}(w)=w^{{\cal C}_1}(w-1)^{{\cal C}_2}(w-w_s)^{{\cal C}_3}(w-w^s_{\infty})^{-1} \mathcal{Y}_s(w),
\end{eqnarray}
we find after a long computation that the angular part is also transformed into the GHE,
\begin{eqnarray}\label{rnc50}
\mathcal{Y}_s^{\prime\prime}(w) +  \bigg(   \frac{\gamma_s}{w} +   \frac{\delta_s}{w  -  1}  +  \frac{\epsilon_s}{w  -  w_s}   \bigg) \mathcal{Y}_s^{\prime}(w)
 +  \frac{\alpha_s\beta_s w-q_s}{w(w  -  1)(w  -  w_s)}  \mathcal{Y}_s(w)=0
,~~~
\end{eqnarray}
with 
\bsubeq\label{apppprefaddd}
\begin{eqnarray}\label{appppref}
&&\gamma_s=2{\cal C}_1, \;\; \quad \delta_s=2{\cal C}_2,  \;\; \quad \epsilon_s=2{\cal C}_3,  \;\; \quad\alpha_s=1, 
\\ % \;\; 
&&\beta_s= 1\mp 2\tilde{\cal B}_4,  \;\; \quad q_s=2{\cal C}_1{\cal C}_3+2{\cal C}_1{\cal C}_2 w_s-{\cal A} w_s,
~~~ \label{apppprefadd}
\end{eqnarray}
and 
\begin{eqnarray}\label{apppppref}
{\cal C}_j=\frac{1}{2} \pm \tilde{\cal B}_j,\;\; \quad  \tilde{\cal B}_j := \frac{m}{\mathcal{P}^{\prime}(x_{j})},\quad  j=1,2,3,4,
\end{eqnarray}
\esubeq
with the following correspondence used here and throughout the rest of our paper
\begin{equation}x_1, x_2, x_3, x_4 \Longleftrightarrow x_{+},x^{\prime}_{+},x^{\prime}_{-},x_{-}.
\end{equation}
Here $\sum_{j=1}^4\tilde{\cal B}_j=0$ and ${\cal A}$ has the same form of $A$ in \eqref{rnc55} with the following replacements: 
$r\mapsto x, z_r\mapsto w_s, z_\infty\mapsto w^s_{\infty}, \tilde{\alpha}\mapsto i \tilde{\alpha} Q, D\mapsto \frac{-4(im)^2}{[\mathcal{P}^{\prime}(x_{-})]^2}(w^s_{\infty})^3$ and $E\mapsto\frac{(im)^2}{[\mathcal{P}^{\prime}(x_{-})]^2} (w^s_{\infty})^4$.
We remark that the parameters in (\ref{rnc50}) satisfy the condition $\gamma_s+\delta_s+\epsilon_s=\alpha_s+\beta_s+1$, as required.

\section{Asymptotic behaviors and Hawking radiation}
\label{rghe}

In this section we determine the asymptotic behaviours of solutions of the radial GHE (\ref{rnc36}) and apply them to analyse Hawking radiation.

\subsection{Asymptotic behaviors}
\label{3a}
 
The radial GHE (\ref{rnc36}) has two linearly independent solutions in the vicinity of $z=0$ 
\cite{hmsn,viqbs,hskd},
\bsubeq\label{rnc555bs10}
\begin{eqnarray}\label{rnc555ea}
\ y_{01}(z)&=&\textrm{HeunG}[a,q_r,\alpha,\beta,\gamma,\delta,\epsilon; z] ,\\ 
\ y_{02}(z)&=&z^{1-\gamma} \,\textrm{HeunG}[a, q_r+(1-\gamma)(\epsilon+a\delta),  1+\beta-\gamma, 1+\alpha-\gamma,2-\gamma,\delta,\epsilon;z],~~~~~~
\end{eqnarray}
\esubeq
where $a=z_r$ and 
$\textrm{HeunG}[\,\ldots,z]$ is the local Heun function  
\begin{eqnarray}\label{heun1}
\textrm{HeunG}[a,q_r,\alpha,\beta,\gamma,\delta,\epsilon; z]=\sum^{\infty}_{n=0}c_nz^n,
\end{eqnarray}
with coefficients $c_n$ defined by the three-term recurrence relation,
\bsubeq\begin{eqnarray}\label{heun2}
&& -q_r c_0+a \gamma c_1=0 ,\\ 
&& P_n c_{n-1}-(Q_n+q_r)c_n+R_n c_{n+1}=0,\; (n \geq 1),
\end{eqnarray}
\esubeq
where 
\bsubeq
\begin{eqnarray}\label{heun3}
 P_n &=&(n-1+\alpha)(n-1+\beta),\\ 
 Q_n &=&n[(n-1+\gamma)(a+1)+a\delta+\epsilon],\\ 
 R_n &=&a(n+1)(n+\gamma).
\end{eqnarray}
\esubeq
The local Heun function \eqref{heun1} is normalized at $z=0$ as
\begin{eqnarray}\label{heun4}
\textrm{HeunG}[a,q_r,\alpha,\beta,\gamma,\delta,\epsilon; 0]=1.
\end{eqnarray}
Asymptotically, when $z \rightarrow 0$, it holds
\bsubeq\label{rncbs1}
\begin{eqnarray}\label{rnc555}
\ y_{01}(z) &\sim& 1+\mathcal{O}(z),\\ 
\ y_{02}(z)&\sim& z^{1-\gamma}(1+\mathcal{O}(z)).
\end{eqnarray}
\esubeq
It then follows from \eqref{rnc22} and \eqref{rerad1} that the asymptotic radial  wave function at the exterior event horizon is given by
\begin{eqnarray}\label{rnc73}
R(r) \sim (r-r_{+})^{\tilde{B}_1} + (r-r_{+})^{- \tilde{B}_1} .
\end{eqnarray}
In deriving \eqref{rnc73}, we have substituted the parameters in \eqref{heun1} by the ones given in \eqref{appref} and \eqref{apprefadd}. The result demonstrates that either sign in the expressions \eqref{appref} and \eqref{apprefadd} can be chosen without affecting the asymptotic behaviour of the radial wave function.

Similarly, at $z=1$, 
%after the substitution of $\tilde{z}=1-z$ into \eqref{rnc36}, we will have its another form,
%\begin{eqnarray}
%\mathcal{Y}^{\prime\prime}(\tilde{z})  +   \bigg(   \frac{\tilde{\gamma}}{\tilde{z}} +   \frac{\tilde{\delta}}{\tilde{z}-1}  +  \frac{\tilde{\epsilon}}{\tilde{z}-\tilde{a}}   \bigg) \mathcal{Y}^{\prime}(\tilde{z})
%  +   \frac{\tilde{\alpha}\tilde{\beta} \tilde{z}-{\tilde{q}_r}}{\tilde{z}(\tilde{z}  -  1)(\tilde{z}  -  \tilde{a})}  \mathcal{Y}(\tilde{z})  = 0, \nonumber
%\end{eqnarray}
%with the parameters substituted as $\tilde{\alpha}=\alpha$, $\tilde{\beta}=\beta$, $\tilde{\gamma}=\delta$, $\tilde{\delta}=\gamma$, $\tilde{\epsilon}=\epsilon$, $\tilde{q}_r=\alpha \beta - q_r $ as well as $\tilde{a}=1-a=1-z_r$.
%With these substitutions into \eqref{rnc555bs10}, 
there will be two linearly independent solutions,
\bsubeq\label{rnc56bs11}
\begin{eqnarray}\label{rnc56}
\ y_{11}(z)&=&\textrm{HeunG}[1-a,\alpha\beta-q_r,\alpha,\beta,\gamma,\delta,\epsilon; 1-z] , \\
\ y_{12}(z)&=&(1-z)^{1-\delta} \textrm{HeunG}[1-a, ((1-a)\gamma+\epsilon)(1-\delta)+\alpha\beta-q_r, \nonumber\\
& &~~~~~~~~~~~~~~~~~~~~~~~~
1+\beta-\delta,1+\alpha-\delta,2-\delta,\gamma,\epsilon; 1-z].~~~~~~
\end{eqnarray}
\esubeq
The local Heun function in \eqref{rnc56} is normalized at $z=1$ in a way similar to \eqref{heun4}. 
Asymptotically, when $z \rightarrow 1$, it holds\bsubeq\label{rncbs2}
\begin{eqnarray}\label{rnc55566}
\ y_{11}(z)&\sim& 1+\mathcal{O}(1-z),\\
\ y_{12}(z)&\sim& (1-z)^{1-\delta}(1+\mathcal{O}(1-z)).
\end{eqnarray}
\esubeq
We thus obtain a similar asymptotic radial wave functions at the acceleration horizon,
\begin{eqnarray}\label{rnc74}
R(r) \sim (r-r^{\p}_{+})^{\tilde{B}_2} + (r-r^{\p}_{+})^{- \tilde{B}_2}.
\end{eqnarray}
Similar to the derivation of \eqref{rnc73}, in deriving this asymptotic radial wave function,  we have also made substitutions of parameters in \eqref{rnc56bs11} by the ones given in \eqref{appref} and \eqref{apprefadd}.

Note that the asymptotic solutions, \eqref{rnc73} and \eqref{rnc74} obtained above, are the same as those obtained via the Damour-Ruffini-Sannan (DRS) method \cite{dr,page2,s}. 

The tortoise coordinate and the surface gravity are defined as follow
\begin{eqnarray}\label{rnc58}
dr_* :=\frac{dr}{\Delta(r)};\;\;\quad \kappa :=\frac{\Delta^{\p}(r_{j})}{2r^2_{j}},\quad j=1,2,3,4.
\end{eqnarray}
This implies that the tortoise coordinate is
\begin{eqnarray}\label{rnc61}
r_*  =\sum^4_j \frac{\ln |r-r_{j}|}{\Delta^{\p}(r)}= \frac{\ln |r-r_{+}|}{2\kappa(r_{+})r^2_{+}} + \frac{\ln |r-r_{-}|}{2\kappa(r_{-})r^2_{-}} + \frac{\ln |r-r^{\p}_{+}|}{2\kappa(r^{\p}_{+})r^{\p 2}_{+}} + \frac{\ln |r-r^{\p}_{-}|}{2\kappa(r^{\p}_{-})r^{\p2}_{-}}.~~~
\end{eqnarray}

In terms of the tortoise coordinate, eq.\,\eqref{rnc5} is written as
\begin{eqnarray}\label{rnc633}
\frac{d^2 R(r)}{dr^2_*}  + \big[\Omega(r)- V_{\textrm{eff}}(r) \big] R(r)= 0,
\end{eqnarray}
where $V_{\textrm{eff}}(r)=-\Delta(r)\big( \frac{1}{6}\Delta^{\p\p}(r)+\lambda \big)$.
In the asymptotic limit $r \rightarrow r_j$, we have
\begin{eqnarray}\label{rnc63}
\frac{d^2 R(r)}{dr^2_*}  + \Omega(r) R(r)= 0.
\end{eqnarray}
Its solutions can be expressed in terms of the parameters of the GHE,
\begin{eqnarray}\label{rnc64}
 R(r)\sim (r-r_j)^{\pm\tilde{B}_j},\;\;j=1,2,
\end{eqnarray}
which are consistent with those from GHE results \eqref{rnc73} and \eqref{rnc74}.

\subsection{Hawking radiation}
\label{3b}
As the first application of the results based on the GHEs, we consider Hawking radiation, which can be interpreted as a scattering problem with wave modes crossing the BH event horizon or the acceleration horizon \cite{hawk4}.

It can be easily checked in the ordering $r^{\p}_{-}< 0\leq r_{-} < r_{+} < r^{\p}_{+}$ that $\kappa(r_{+})>0$ and $\kappa(r^{\p}_{+})<0$. Thus, from the results in the previous section,
the spatial-dependent ingoing and outgoing waves have the form,
\begin{eqnarray}\label{ppc3}
R_{\textrm{in/out}}(r>r_j)=(r-r_j)^{\pm \frac{ i J(r_{j})}{2 \kappa (r_j) r^2_j}}.
\end{eqnarray}
As per the different signs of $\textrm{Re}[J]$, there will be two branches.

For $\textrm{Re}[J]>0$, 
we have the following ingoing/outgoing solutions around different horizons.
Around the event horizon $(r - r_{+}) \rightarrow 0^+$,
\bsubeq\begin{eqnarray}\label{ppc4}
\ R_{\textrm{in}}(r>r_+)&=&(r-r_+)^{-\frac{i J(r_{+})}{2 \kappa (r_+) r^2_+}},\\
\ R_{\textrm{out}}(r>r_+)&=&(r-r_+)^{\frac{ i J(r_+)}{2 \kappa (r_+) r^2_+}},
\end{eqnarray}
\esubeq
and around the acceleration horizon $(r - r^{\p}_{+}) \rightarrow 0^+$,
\bsubeq\begin{eqnarray}\label{ppc5}
\ R_{\textrm{in}}(r>r^{\p}_{+})&=&(r-r^{\p}_{+})^{\frac{i J(r^{\p}_{+})}{2 \kappa (r^{\p}_{+}) r^{\p 2}_{+}}},\\ 
\ R_{\textrm{out}}(r>r^{\p}_{+})&=&(r-r^{\p}_{+})^{-\frac{i J(r^{\p}_{+})}{2 \kappa (r^{\p}_{+}) r^{\p 2}_{+}}}.
\end{eqnarray}
\esubeq

Considering the time-dependent damping properties, focusing on the outgoing waves, we have
\bsubeq\begin{eqnarray}\label{ppc6}
\ R_{\textrm{out}}(t,r>r_+)&=&e^{-i \omega t}(r-r_+)^{\frac{ i J(r_+)}{2 \kappa (r_+) r^2_+}},  \\
\ R_{\textrm{out}}(t,r>r^{\p}_{+})&=&e^{-i \omega t}(r-r^{\p}_{+})^{-\frac{i J(r^{\p}_{+})}{2 \kappa (r^{\p}_{+}) r^{\p 2}_{+}}}.\label{ppc6b}
\end{eqnarray}
\esubeq
As a result, the outgoing wave from the event horizon is \eqref{ppc6}.

This wave can be analytically continued inside the event horizon ($r_-<r< r_+$) by the lower half complex $r$-plane around the unit circle $r=r_+-i0$: $r-r_+ \rightarrow (r_+-r)e^{-i\pi}$ \cite{sqw},
\begin{eqnarray}\label{rncadd1}
 R_{\textrm{out}}^c(t,r<r_+)&=&e^{-i \omega t}[(r_+-r)e^{-i\pi}]^{\frac{ i J(r_{+})}{2 \kappa (r_+) r^2_+}}  
 \nonumber \\
&=&e^{-i \omega t}(r_+-r)^{\frac{ i J(r_{+})}{2 \kappa (r_+) r^2_+}} (e^{-i\pi})^{\frac{ i [\textrm{Re}[J(r_{+})]+i \textrm{Im} [J(r_{+})]}{2 \kappa (r_+) r^2_+}} \nonumber \\
&=&e^{-i \omega t}\bigg[(r_+-r)^{\frac{ i J(r_{+})}{2 \kappa (r_+) r^2_+}}\bigg]  \bigg[e^{\frac{ i \pi  \textrm{Im} [J(r_{+})]}{2 \kappa (r_+) r^2_+}}\bigg] \bigg[e^{\frac{ \pi \textrm{Re}[J(r_{+})]}{2 \kappa (r_+) r^2_+}}\bigg], 
\end{eqnarray}
where $R_{\textrm{out}}^c(t, r<r_+)$ denotes the analytically continued outgoing wave.

After analytic continuation, we can get the scattering probability and outgoing energy density for Hawking radiation:
\bsubeq\label{rncf6}
\begin{eqnarray}\label{rnc66}
\Gamma(J)_{r_+}= \bigg|\frac{R_{\textrm{out}}(t,r>r_+)}{R_{\textrm{out}}^c(t,r<r_+)} \bigg|^2 = e^{- \frac{\pi \textrm{Re}[J(r_+)]}{\kappa (r_+) r^2_+}},
\end{eqnarray}
\begin{eqnarray}\label{rnc666}
N(J)_{r_+}= \frac{\Gamma_J}{1-\Gamma_J}= \frac{1}{e^{\frac{\hbar \textrm{Re}[J(r_+)]}{k_B T_+ r^2_+}}-1},
\end{eqnarray}
\esubeq
where $k_B$ is the Boltzmann constant and $T_+$ the Hawking-Unruh temperature at the event horizon: $k_B T_+=\frac{\hbar \kappa (r_+) }{\pi}$ \cite{miladd}\cite{miladd2}. 
When the $qQ$-dependent term can be ignored, and the frequency $\omega$ is real, i.e.  $\textrm{Re}J(r_+) \rightarrow \omega r^2_+$, the results in \eqref{rncf6} reduce to those in \cite{sqw,hsvc2,hsvc4} for a spherically symmetric background,
\begin{eqnarray}\label{rnc6886}
\Gamma(J)_{r_+} \rightarrow e^{\frac{-\pi \omega }{\kappa (r_+)}};\;\;\; N(J)_{r_+} \rightarrow N(\omega)_{r_+} =\frac{1}{e^{\frac{\hbar \omega}{k_B T_+}-1}}.
\end{eqnarray}

Similarly, we can also get the Hawking radiation for the acceleration horizon
\bsubeq\begin{eqnarray}
\Gamma(J)_{r^{\p}_{+}} &=& e^{ \frac{\pi \textrm{Re}[J(r^{\p}_{+})]}{\kappa (r^{\p}_{+}) r^{\p 2}_+}},\label{rnc66add}\\
N(J)_{r^{\p}_{+}}&=& \frac{1}{e^{-\frac{\hbar \textrm{Re}[J(r^{\p}_{+})]}{k_B T^{\prime}_+ r^{\p 2}_+}}-1},\label{rnc666add}
\end{eqnarray}
\esubeq
where $T^{\prime}_+$ is the Hawking-Unruh temperature at the acceleration horizon: $k_B T^{\prime}_+=\frac{\hbar \kappa (r^{\p}_{+}) }{\pi}$.

For $\textrm{Re}[J]<0$, after the same procedure as above, for the event horizon, we obtain
\bsubeq
\begin{eqnarray}
\Gamma(J)_{r_+}= e^{ \frac{\pi \textrm{Re}[J(r_+)]}{\kappa (r_+) r^2_+}},
\end{eqnarray}
\begin{eqnarray}
N(J)_{r_+}= \frac{\Gamma_J}{1-\Gamma_J}= \frac{1}{e^{-\frac{\hbar \textrm{Re}[J(r_+)]}{k_B \tilde{T}_+ r^2_+}}-1},
\end{eqnarray}
\esubeq
with $k_B \tilde{T}_+=-\frac{\hbar \kappa (r_+) }{\pi}$. 
At the acceleration horizon, it holds
\bsubeq\begin{eqnarray}
\Gamma(J)_{r^{\p}_{+}} &=& e^{ -\frac{\pi \textrm{Re}[J(r^{\p}_{+})]}{\kappa (r^{\p}_{+}) r^{\p 2}_+}},\\
N(J)_{r^{\p}_{+}}&=& \frac{1}{e^{\frac{\hbar \textrm{Re}[J(r^{\p}_{+})]}{k_B \tilde{T}^{\prime}_+ r^{\p 2}_+}}-1},
\end{eqnarray}
\esubeq
with $k_B \tilde{T}^{\prime}_+=-\frac{\hbar \kappa (r^{\p}_{+}) }{\pi}$.

\section{Boundary conditions for Quasinormal modes and superradiance at asymptotic limits}
\label{bound}

Asymptotically, the boundary conditions of the wave functions can generally be characterized by the tortoise coordinates $r_* \rightarrow \pm \infty$ and $x_* \rightarrow \pm \infty$ ($x \rightarrow \mp 1$) for the radial and angular parts, respectively \cite{ccmetric,ebvcas}.

In the following, we discuss the asymptotic behaviors of the GHE solutions in general without considering any specific boundary or regularity conditions. We will then discuss the boundary conditions for two applications, QNMs and superradiance.

\subsection{Coefficients of asymptotic behaviors in the tortoise coordinate}
\label{3c}

We first determine the proportionality coefficients in the asymptotic wave functions \eqref{rnc73}, \eqref{rnc74} and \eqref{rnc64}. 

From the definition of tortoise coordinate in eq.\,\eqref{rnc61}, asymptotically when $r \rightarrow r_j$, we have
\begin{eqnarray}\label{tornc1}
r_* \sim \sum^4_j \frac{\ln |r-r_{j}|}{\Delta^{\p}(r_{j})}+ d_j,
\end{eqnarray}
where $d_j$ are constant coefficients defined by 
\begin{equation}d_j :=\sum_{k\neq j} \frac{\ln |r_j-r_{k}|}{\Delta^{\p}(r_{k})},\;\quad j,k=1,2,3,4.
\end{equation}
Solving \eqref{tornc1} gives,
\begin{eqnarray}\label{tornc2}
r-r_{j} \sim e^{-d_j \Delta^{\p}(r_{j})} e^{ \Delta^{\p}(r_{j}) r_*}.
\end{eqnarray}
Then from \eqref{rerad1}, \eqref{rnc22}, \eqref{rncbs1} and \eqref{rncbs2}, we can determine the asymptotic wave solutions in terms of the tortoise coordinate $r_*$ as follows.

Around $z=0$, we have,
\bsubeq\label{torncbs3}
\begin{eqnarray}\label{tornc3ad1}
\ R_{01}(r\rightarrow r_+)\sim  (r- r_+)^{\pm \tilde{B}_1},\\
\ R_{02}(r\rightarrow r_+)\sim  (r- r_+)^{\mp \tilde{B}_1},\label{tornc3ad1b}
\end{eqnarray}
\esubeq
or in tortoise coordinate, $r_* \rightarrow -\infty$, 
\bsubeq\label{torncbs4}
\begin{eqnarray}\label{tornc3}
\  R_{01}(r_* \rightarrow -\infty) &\sim&  A_{01}^{(\mp)} e^{\pm i J(r_{+})r_*}, \\
\  R_{02}(r_* \rightarrow -\infty) &\sim&  A_{02}^{(\pm)} e^{\mp i J(r_{+})r_*},\label{tornc3b}
\end{eqnarray}
\esubeq
where the constant coefficients are
\bsubeq
\begin{eqnarray}\label{tornc4}
\       A_{01}^{(\mp)}&=&\frac{(-1)^{C_2}(-z_r)^{C_3}(-z_{\infty})^{-1}}{\sqrt{-\tilde{\alpha}^2 \prod_{k\neq 1}(r_{+}-r_k)}}\, e^{\mp i d_1 J(r_{+})}, \\
\       A_{02}^{(\pm)}&=&\frac{(-1)^{C_2}(-z_r)^{C_3}(-z_{\infty})^{-1}}{\sqrt{-\tilde{\alpha}^2 \prod_{k\neq 1}(r_{+}-r_k)}}\, e^{\pm i d_1 J(r_{+})}.
\end{eqnarray}
\esubeq

Around $z=1$, we get,
\bsubeq\label{torncbs5}
\begin{eqnarray}\label{tornc3ad11}
\ R_{11}(r\rightarrow r^{\p}_{+})\sim  (r- r^{\p}_{+})^{\pm \tilde{B}_2},\\
\ R_{12}(r\rightarrow r^{\p}_{+})\sim  (r- r^{\p}_{+})^{\mp \tilde{B}_2},\label{tornc3ad11b}
\end{eqnarray}
\esubeq
or in tortoise coordinate, or $r_* \rightarrow +\infty$, 
\bsubeq\label{torncbs6}
\begin{eqnarray}\label{tornc5}
\ R_{11}(r_* \rightarrow +\infty) \sim A_{11}^{(\mp)}e^{\pm i J( r^{\p}_{+}) r_*}, \\
\ R_{12}(r_* \rightarrow +\infty) \sim A_{12}^{(\pm)}e^{\mp i J( r^{\p}_{+}) r_*}, \label{tornc5b}
\end{eqnarray}
\esubeq
with the following constant coefficients,
\bsubeq
\begin{eqnarray}\label{tornc6}
\       A_{11}^{(\mp)}&=&\frac{(1)^{C_1}(1   -   z_r)^{C_3}(1   -   z_{\infty})^{-1}}{\sqrt{-\tilde{\alpha}^2 \prod_{k\neq 2}(r^{\p}_{+}  -   r_k)}}\, e^{\mp i d_2 J(r^{\p}_{+})}, \\
\       A_{12}^{(\pm)}&=&\frac{(1)^{C_1}(1   -   z_r)^{C_3}(1   -   z_{\infty})^{-1}}{\sqrt{-\tilde{\alpha}^2 \prod_{k\neq 2}(r^{\p}_{+}  -   r_k)}}\, e^{\pm i d_2 J(r^{\p}_{+})}.
\end{eqnarray}
\esubeq
Similarly, we can determine the coefficients for the angular solutions from \eqref{reang1}, \eqref{angrnc22}, \eqref{rncbs1} and \eqref{rncbs2} as follows.

From \eqref{rnc4}, after making a similar variable change, one obtain an equation similar to \eqref{rnc5}, 
\begin{eqnarray}\label{rnc5add}
\mathcal{X}^{\p\p}(x) +\frac{\mathcal{P}^{\p}(x)}{\mathcal{P}(x)} \mathcal{X}^{\p}(x)
+\frac{1}{\mathcal{P}(x)}\bigg[\frac{-m^2}{\mathcal{P}(x)}  +   \frac{1}{6}\mathcal{P}^{\p\p}(x)  +   \lambda \bigg] \mathcal{X}(x)= 0,
\end{eqnarray}
Let $w_*$ denote the tortoise coordinate for the angular equation, which is defined as,
\begin{eqnarray}\label{rnc58add}
dw_* :=\frac{dx}{\mathcal{P}(x)}; \;\; w_* \sim \sum^4_j \frac{\ln |x-x_{j}|}{\mathcal{P}^{\p}(x)}+ d^s_j,
\end{eqnarray}
where $d^s_j$ are constant coefficients defined by 
\begin{equation}\label{notean2}
d^s_j :=\sum_{k\neq j} \frac{\ln |x_j-x_{k}|}{\mathcal{P}^{\p}(x_{k})},\;\quad j,k=1,2,3,4.
\end{equation}
Solving \eqref{rnc58add} gives,
\begin{eqnarray}\label{notean3}
x-x_{j} \sim e^{-d^s_j \mathcal{P}^{\p}(x_{j})} e^{ \mathcal{P}^{\p}(x_{j}) w_*}.
\end{eqnarray}
In terms of $w_*$, equation \eqref{rnc5add} takes a form analogue to \eqref{rnc633},
\begin{eqnarray}\label{rnc633add}
\frac{d^2 \mathcal{X}(x)}{dw^2_*}  + \big[-m^2- V_{\textrm{eff}}(x) \big] \mathcal{X}(x)= 0,
\end{eqnarray}
where $V_{\textrm{eff}}(x)=-\mathcal{P}(x)\big( \frac{1}{6}\mathcal{P}^{\p\p}(x)+\lambda \big)$.
In the asymptotic limit $x \rightarrow x_j$, we have
\begin{eqnarray}\label{rnc63333}
\frac{d^2 \mathcal{X}(x)}{dw^2_*}  -m^2 \mathcal{X}(x)= 0.
\end{eqnarray}
Its boundary conditions are
\begin{eqnarray}\label{rnc633addd}
\mathcal{X}(x) \sim e^{ \pm m w_*}, 
\end{eqnarray}
In terms of the parameters of the GHE, this become
\begin{eqnarray}\label{rnc64444}
 \mathcal{X}(x) \sim (x-x_j)^{\frac{1}{2}+ {\cal C}_j}, (x-x_j)^{\frac{3}{2}- {\cal C}_j}\;\;j=1,2.
\end{eqnarray}

In more detail, around $w=0$ ($x\rightarrow x_+$), it holds
\bsubeq\label{tornc11-0}
\begin{eqnarray}\label{tornc11}
\  \mathcal{X}_{01}(x\rightarrow x_+) &\sim&  (x-x_+)^{(1 \pm \tilde{B}^s_1)}, \\
\  \mathcal{X}_{02}(x \rightarrow x_+) &\sim&  (x-x_+)^{(1 \mp \tilde{B}^s_1)}.\label{tornc11b}
\end{eqnarray}
\esubeq
In terms of $w_*$, we have 
\bsubeq\label{tornc7}
\begin{eqnarray}
\  \mathcal{X}_{01}(w_* \rightarrow +\infty) &\sim&  A^{(\pm)s}_{01} e^{(\mathcal{P}^{\prime}(x_{+}) \pm m) w_*}, \\
\  \mathcal{X}_{02}(w_* \rightarrow +\infty) &\sim&  A^{(\mp)s}_{02} e^{(\mathcal{P}^{\prime}(x_{+}) \mp m) w_*},
\end{eqnarray}
\esubeq
with $\mathcal{P}^{\prime}(x_{+}) =-2(1+2\tilde{\alpha} M +\tilde{\alpha}^2 Q^2)$ and 
\bsubeq\label{tornc8}
\begin{eqnarray}
\       A^{(\pm)s}_{01}  &=&  (-1)^{C^s_2}(-w_s)^{C^s_3}(-w^s_{\infty})^{-1}\sqrt{\tilde{\alpha}^2 Q^2 \prod_{k\neq 1}(x_{+}  -  x_k)}~ e^{- d^s_1 \mathcal{P}^{\prime}(x_{+})(1\pm \tilde{B}^s_1)},
\\
\       A^{(\mp)s}_{02}  &=&  (-1)^{C^s_2}(-w_s)^{C^s_3}(-w^s_{\infty})^{-1}\sqrt{\tilde{\alpha}^2 Q^2 \prod_{k\neq 1}(x_{+}  -  x_k)} ~e^{- d^s_1 \mathcal{P}^{\prime}(x_{+})(1\mp \tilde{B}^s_1)}.
\end{eqnarray}
\esubeq

Around $w=1$ ($x\rightarrow x^{\p}_+$), it holds
\bsubeq\label{torncbs7}
\begin{eqnarray}\label{tornc12}
\  \mathcal{X}_{11}(x \rightarrow x^{\p}_+) &\sim&  (x-x^{\p}_+)^{(1 \pm \tilde{B}^s_2)},\\
\  \mathcal{X}_{12}(x \rightarrow x^{\p}_+) &\sim&  (x-x^{\p}_+)^{(1 \mp \tilde{B}^s_2)}.\label{tornc12b}
\end{eqnarray}
\esubeq
In terms of $w_*$, they are
\bsubeq\label{tornc9}
\begin{eqnarray}
\  \mathcal{X}_{11}(w_* \rightarrow -\infty) &\sim&  A^{(\pm)s}_{11} e^{(\mathcal{P}^{\prime}(x^{\p}_+) \pm m) w_*}, \\
\  \mathcal{X}_{12}(w_* \rightarrow -\infty) &\sim&  A^{(\mp)s}_{12} e^{(\mathcal{P}^{\prime}(x^{\p}_+) \mp m) w_*}
,
\end{eqnarray}
\esubeq
with $\mathcal{P}^{\prime}(x^{\p}_+) =2(1-2\tilde{\alpha} M +\tilde{\alpha}^2 Q^2)$ and
\bsubeq
\begin{eqnarray}\label{tornc10}
\       A^{(\pm)s}_{11}  &=&  (1)^{C^s_1}(1  -  w_s)^{C^s_3}(1  -   w^s_{\infty})^{-1}\sqrt{\tilde{\alpha}^2 Q^2 \prod_{k\neq 2}(x^{\p}_+  -  x_k)} ~e^{- d^s_2 \mathcal{P}^{\prime}(x^{\p}_+)(1\pm \tilde{B}^s_2)},\\
\       A^{(\mp)s}_{12}  &=&  (1)^{C^s_1}(1  -  w_s)^{C^s_3}(1  -   w^s_{\infty})^{-1}\sqrt{\tilde{\alpha}^2 Q^2 \prod_{k\neq 2}(x^{\p}_+  -  x_k)}~ e^{- d^s_2 \mathcal{P}^{\prime}(x^{\p}_+)(1\mp \tilde{B}^s_2)}.
\end{eqnarray}
\esubeq
Comparing these equations with \eqref{tornc7}, \eqref{tornc9} and \eqref{rnc633addd},  we see that the exact results are different from the asymptotic solutions. 

\subsection{Applications to QNMs}
\label{qnmapp}

When a wave propagates near or crosses the horizon of a BH, the extreme gravitational time dilation causes a significant alteration in the observed frequency. This effect usually manifests as a discrete set of complex frequencies, the so-called QNMs, which characterize the decay or growth of the wave. 
The real part of the QNM describes the damped oscillation, and the imaginary part the exponential decay, provided it is negative, while a positive imaginary part indicates a dynamical instability \cite{ajo}.

QNMs can generally be understood as an eigenvalue problem of the wave equation in the BH background under the following boundary conditions, characterized by the frequencies $\omega$ \cite{ebvcas,edwl,nacjh,ebkdk}.
More precisely, there are only ingoing waves at the event horizon, while the ingoing modes are discarded at the acceleration horizon because they are unobservable beyond this horizon.

In view of \eqref{torncbs3}, the outgoing and ingoing wave modes at the event horizon are  (provided that Re$J(r_+)> 0$),
\bsubeq
\begin{eqnarray}
\  \textrm{outgoing}: \;\; R_{01}(r\rightarrow r_+)\sim  (r- r_+)^{- \tilde{B}_1},\\
\  \textrm{ingoing}: \;\; R_{02}(r\rightarrow r_+)\sim  (r- r_+)^{+ \tilde{B}_1},
\end{eqnarray}
\esubeq
or from \eqref{torncbs4},
\bsubeq\label{torncbs8}
\begin{eqnarray}\label{tornc13}
\  \textrm{outgoing}: \;\; R_{01}(r_* \rightarrow -\infty) &\sim & A_{01}^{(+)} e^{- i J(r_{+})r_*},\\
\  \textrm{ingoing}: \;\; R_{02}(r_* \rightarrow -\infty) &\sim & A_{02}^{(-)} e^{+ i J(r_{+})r_*}, \label{tornc13b}
\end{eqnarray}
\esubeq
since $r_* \rightarrow -\infty$ at $r \rightarrow r_+$.

Similarly, from equation \eqref{torncbs5}, we have outgoing and ingoing wave modes (provided that Re$J(r^{\p}_+)< 0$),
\bsubeq
\begin{eqnarray}
\ \textrm{outgoing}: \;\; R_{11}(r\rightarrow r^{\p}_+)\sim  (r- r^{\p}_+)^{- \tilde{B}_2},\\
\  \textrm{ingoing}: \;\; R_{12}(r\rightarrow r^{\p}_+)\sim  (r- r^{\p}_+)^{+ \tilde{B}_2},
\end{eqnarray}
\esubeq
or from \eqref{torncbs6},
\bsubeq\label{torncbs9}
\begin{eqnarray}\label{tornc15}
\  \textrm{outgoing}: \;\; R_{11}(r_* \rightarrow +\infty) &\sim&  A_{11}^{(+)} e^{- i J(r^{\p}_+)r_*}, \\
\  \textrm{ingoing}: \;\; R_{12}(r_* \rightarrow +\infty) &\sim&  A_{12}^{(-)} e^{+ i J(r^{\p}_+)r_*}, \label{tornc15b}
\end{eqnarray}
\esubeq
since $r_* \rightarrow +\infty$ at $r \rightarrow r^{\p}_+$. It follows that
the wave functions under the boundary conditions to determine the QNMs are 
\begin{eqnarray}\label{tornc17}
R(r) \sim \left\{
\begin{array}{ll} 
\  R_{02}(r), \quad r \rightarrow r_+, \\ \\
\  R_{11}(r), \quad r \rightarrow r^{\p}_+.
\end{array}
\right.
\end{eqnarray}

Based on \eqref{tornc17}, we can use the standard matching procedure to derive the QNMs \cite{cccmetric,ccmetric,ccccmetric}.
The logic of this method is similar to that for studying quasibound modes (QBMs): Solving the radial GHE in two different asymptotic regions and matching the solutions in their overlapped region (see, e.g. \cite{hskd,viqbs} and references therein).

To apply this procedure, we should determine the solutions of the angular GHE and their regularity conditions. 
It can be shown that, 
to ensure regularity, we need to impose $|m|<2$.
From \eqref{tornc11-0} at $x\rightarrow x_+$ ($w_* \rightarrow +\infty$), we have
\bsubeq\label{tornc18add}
\begin{eqnarray}\label{tornc18}
\  \mathcal{X}_{01}(x\rightarrow x_+) \sim A^{(-)s}_{01} (x-x_+)^{1- \tilde{\cal B}_1},~~~ |m|<2, \\
\  \mathcal{X}_{02}(x\rightarrow x_+) \sim A^{(-)s}_{02} (x-x_+)^{1+ \tilde{\cal B}_1},~~~ |m|<2,\label{tornc18-2}
\end{eqnarray}
\esubeq
Similarly from \eqref{torncbs7} at $x\rightarrow x^{\p}_+$ ($w_* \rightarrow -\infty$), we get
\bsubeq\label{tornc20add}
\begin{eqnarray}\label{tornc20}
\  \mathcal{X}_{11}(x\rightarrow x^{\p}_+) \sim A^{(-)s}_{11} (x-x^{\p}_+)^{1- \tilde{\cal B}_2}, ~~~|m|<2,\\
\  \mathcal{X}_{12}(x\rightarrow x^{\p}_+) \sim A^{(-)s}_{12} (x-x^{\p}_+)^{1+ \tilde{\cal B}_2}, ~~~|m|<2.\label{tornc20b}
\end{eqnarray}
\esubeq

\subsection{Applications to superradiance}
\label{sup}

Superradiance denotes the interaction between the wave functions and the BH's rotations.
An incident wave from the accelerating horizon, scattering off the BH, will be partially reflected backward the acceleration horizon and partially transmitted through the electromagnetic potential barrier and into the event horizon \cite{ccmetric}. 
We remark that unlike rotating BHs, which utilize the ergoregion to extract energy via Penrose process \cite{rprm,jdb,kdun,rbvcpp} (see also Chapter 6.6-6.7 of \cite{carroll}), the superradiance here is generated by the electromagetic potential barrier between the event horizon ($r_{+}$) and the acceleration horizon ($r_{+}^{\p}$).

The boundary conditions for this kind of superradiance contain the ingoing wave modes at the event horizon and the transmitting wave modes (outgoing and ingoing) across the acceleration horizon \cite{ccmetric}.
From \eqref{torncbs8} and \eqref{torncbs9}, 
\begin{eqnarray}\label{tornc22}
R(r) \sim \left\{
\begin{array}{ll} 
\  R_{02}(r), \quad r \rightarrow r_+, \\ \\
\  R_{11}(r)+R_{12}(r), \quad r \rightarrow r^{\p}_+.
\end{array}
\right.
\end{eqnarray}

To obtain the reflection and transmission amplitudes, we write
\begin{eqnarray}\label{tornc222222}
R(r) \sim \left\{
\begin{array}{ll} 
\  \mathcal{T} e^{+ i J(r_{+})r_*}, \quad r \rightarrow r_+, \\ \\
\  \mathcal{I} e^{+ i J(r^{\p}_+)r_*}+ \mathcal{R} e^{- i J(r^{\p}_+)r_*}, \quad r \rightarrow r^{\p}_+,
\end{array}
\right.
\end{eqnarray}
where $\mathcal{T}$ is the transmission wave into the event horizon, $\mathcal{I}$ the incident wave into the acceleration horizon and $\mathcal{R}$ the reflection wave away from the acceleration horizon.
Then, the Wronskians for these waves and their linearly independent solutions are 
\bsubeq
\begin{eqnarray}\label{torn2222}
 W [\mathcal{T} e^{+ i J(r_{+})r_*}, \mathcal{T}^* e^{- i J(r_{+})r_*}]&=&|\mathcal{T}|^2 [-2i J(r_{+})],\\
 W [\mathcal{I} e^{+ i J(r^{\p}_+)r_*}, \mathcal{I}^* e^{- i J(r^{\p}_+)r_*}]&=&|\mathcal{I}|^2 [-2i J(r^{\p}_+)], \\
 W [\mathcal{R} e^{- i J(r^{\p}_+)r_*}, \mathcal{R}^* e^{ +i J(r^{\p}_+)r_*}]&=&|\mathcal{R}|^2 [+2i J(r^{\p}_+)].~~~~~~
\end{eqnarray}
\esubeq
They must coincide at both  boundaries, i.e.
\begin{eqnarray}\label{tor2222}
|\mathcal{T}|^2 [-2i J(r_{+})]=|\mathcal{I}|^2 [-2i J(r^{\p}_+)]+|\mathcal{R}|^2 [2i J(r^{\p}_+)].
\end{eqnarray}
This yields
\begin{eqnarray}\label{tor2222333}
|\mathcal{R}|^2 =|\mathcal{I}|^2 -\frac{J(r_{+})}{J(r^{\p}_+)} |\mathcal{T}|^2 
.
\end{eqnarray}

As long as $\frac{J(r_{+})}{J(r^{\p}_+)}<0$, the reflection amplitude will be larger than the incident amplitude, which is the indication that there is superradiance.
This agrees with the result in \cite{ccmetric}.
However, we still need to settle the parameters to calculate the exact superradiant amplifications.
This can be achieved using the exact solutions.

\section{Exact GHE and applications to Quasinormal modes and superradiance}
\label{gheeeb}

In the previous section (Sec.\ref{qnmapp}), we studied the asymptotic behaviours of the radial wave functions in the respective convergence region around $z=0$ and $z=1$ and in their overlapped convergence region by matching the solutions. 

In this section, we explore the exact connection-coefficient formalism proposed in \cite{hmsn}, which provides an effective approach to many of the scattering-related problems. Specifically, the connection coefficients enable us to establish an overlapped convergence disk, in which we can determine the exact behaviors of the radial solutions at an arbitrary point in the disk without analytic continuation.

Let $C_{11}, C_{12}, C_{21}, C_{22}$ denote the connection coefficients connecting the two exact solutions in \eqref{rnc555bs10} (of convergence at $z=0$) and \eqref{rnc56bs11} (which is convergent at $z=1$). Then, by definition, in the overlapped convergence region, the two solutions are related by  
\bsubeq
\begin{eqnarray}\label{rnc57}
\  y_{01}(z)&=& C_{11}y_{11}(z)+C_{12}y_{12}(z),\\
\  y_{02}(z)&=& C_{21}y_{11}(z)+C_{22}y_{12}(z).
\end{eqnarray}
\esubeq

From Cramer's rule \cite{erpb}, we can obtain the connection coefficients around $z=0$,
\bsubeq
\begin{eqnarray}\label{rnc76}
 C_{11}=\frac{W[y_{01},y_{12}]}{W[y_{11},y_{12}]}&,&\qquad C_{12}=\frac{W[y_{01},y_{11}]}{W[y_{12},y_{11}]},  \\
C_{21}=\frac{W[y_{02},y_{12}]}{W[y_{11},y_{12}]}&,&\qquad C_{22}=\frac{W[y_{02},y_{11}]}{W[y_{12},y_{11}]},
\end{eqnarray}
\esubeq

where $W$ is the Wronskian,
 \begin{equation}
 W[f_1(z),f_2(z)]=\bigg| \begin{array}{cc}
f_1(z)& f_2(z)\\
f^{\p}_1(z)& f^{\p}_2(z)\\
\end{array} \bigg|,
\end{equation}
in which the prime denotes the derivative with respect to the variable $z$.

Around $z=1$ \cite{hmsn,snhm}, they are similarly given by
\bsubeq
\begin{eqnarray}\label{rnc57dij}
\  y_{11}(z)&=& D_{11}y_{01}(z)+D_{12}y_{02}(z),\\
\  y_{12}(z)&=& D_{21}y_{01}(z)+D_{22}y_{02}(z),
\end{eqnarray}
\esubeq 
where
\bsubeq
\begin{eqnarray}\label{rnc77}
D_{11}=\frac{W[y_{11},y_{02}]}{W[y_{01},y_{02}]}&,&\qquad D_{12}=\frac{W[y_{11},y_{01}]}{W[y_{02},y_{01}]},   \\
D_{21}=\frac{W[y_{12},y_{02}]}{W[y_{01},y_{02}]}&,&\qquad D_{22}=\frac{W[y_{12},y_{01}]}{W[y_{02},y_{01}]}.
\end{eqnarray}
\esubeq

Consider the asymptotic behaviors of the wave functions in terms of the connection coefficients. Within the context of conformal diagrams \cite{mgigm}, we can distinguish the wave functions in the four parts with different boundary conditions as follows \cite{hmsn,fwdpcl}.

\begin{itemize}

\item $\mathcal{R}_{\textrm{in}}(r)$: around $z=0$ or $r=r_{+}$, with a reference horizon between $z=0$ and $z=1$, no outgoing waves from $z=0$;

Corresponding to $r_* \rightarrow -\infty$, only ingoing modes should be present for the event horizon.

\item $\mathcal{R}_{\textrm{up}}(r)$: around $z=1$ or $r=r^{\p}_+$, with a reference horizon (or potential barrier) between $z=0$ and $z=1$, no ingoing waves from $z=1$;

Corresponding to $r_* \rightarrow +\infty$, only outgoing modes should be present for the acceleration horizon.

\item $\mathcal{R}_{\textrm{down}}(r)=\mathcal{R}^*_{\textrm{up}}(r)$.

\item $\mathcal{R}_{\textrm{out}}(r)=\mathcal{R}^*_{\textrm{in}}(r)$.
\end{itemize}

Their relations with the GHE solutions  are \cite{hmsn,fwdpcl,snhm}
\begin{eqnarray}\label{qmbb78}
\mathcal{R}_{\textrm{in}}(r) = \left\{
\begin{array}{ll} 
\ R_{02}(r), \;\; r \rightarrow r_{+};\\ \\
\ C_{21}R_{11}(r)+C_{22}R_{12}(r), \;\; r \rightarrow r^{\p}_+,
\end{array}
\right.
\end{eqnarray}
\begin{eqnarray}\label{rnc79}
\mathcal{R}_{\textrm{up}}(r) = \left\{
\begin{array}{ll} 
\ D_{11}R_{01}(r)+D_{12}R_{02}(r), \;\; r \rightarrow r_{+};\\ \\
\ R_{11}(r), \;\; r \rightarrow r^{\p}_+,
\end{array}
\right.
\end{eqnarray}
where 
\begin{equation}
R_{Ii}=\Delta^{-\frac{1}{2}}(r)z^{\frac{\gamma}{2}}(z-1)^{\frac{\delta}{2}}(z-z_r)^{\frac{\epsilon}{2}} (z-z_{\infty})^{-1}y_{Ii}(z),
\end{equation}
with $I=0,1$ and $i=1,2$.
Similar relations can be found for the angular wave solutions $\mathcal{X}^s_{Ii}$.

Comparing \eqref{qmbb78} with \eqref{tornc17}, it can be seen that in terms of the connection coefficients, the QNMs are obtained from \eqref{qmbb78} by setting $C_{22}=0$, or equivalently from \eqref{rnc79} by requiring $D_{11}=0$. 
For the superradiance, in view of \eqref{tornc22}, we find that in terms of the connection coefficient formalism, one should look at equation \eqref{qmbb78}.  Comparing with \eqref{tor2222333}, we obtain
\begin{eqnarray}\label{tor222233}
|C_{21}|^2|A_{11}|^2 =|C_{22}|^2|A_{12}|^2 -\frac{J(r_{+})}{J(r^{\p}_+)} |A_{02}|^2 .
\end{eqnarray}

One defines the amplification factor as \cite{rbvcpp,ccmetric}
\begin{eqnarray}\label{amp1}
Z_{0m}=\frac{|\mathcal{R}|^2}{|\mathcal{I}|^2}-1,
\end{eqnarray}
where 
``0'' denotes the scalar field with spin weight $s=0$.
Then, the amplification factor for the superradiance is \cite{hmsn,snhm}  \begin{eqnarray}\label{tornc23}
Z_{0m}=- \frac{J(r_+)}{J(r^{\p}_+)} \bigg| \frac{\prod_{k\neq 2}(r^{\p}_+-r_k)}{\prod_{k\neq 1}(r_{+}-r_k)]} \bigg|
\bigg(\frac{1-z_{\infty}}{z_{\infty}} \bigg)^2 \frac{1}{|C_{22}|^2},
\end{eqnarray}
which can be simplified to 
\begin{eqnarray}\label{tornc23add}
Z_{0m}= - \frac{\tilde{B}_1}{\tilde{B}_2} 
\bigg(\frac{1-z_{\infty}}{z_{\infty}} \bigg)^2 \frac{1}{|C_{22}|^2}.
\end{eqnarray}
Here we have used $\prod_{k\neq j}(r^{\prime}_{j}-r_k)=\Delta^{\prime}(r_{j})$ and $\tilde{B}_j :=  i  \frac{J(r_{j})}{\Delta^{\prime}(r_{j})}$, with $j,k=1,2,3,4$.
We can then see that superradiance exists in \eqref{tornc23}, provided that $\frac{ J(r_+)}{J(r^{\p}_+)}<0$. This agrees with the result in \cite{hmsn}.

\section{Conclusions and Discussions}
\label{sdos}

We have shown that the radial and angular ODEs of the conformally invariant wave equation for the massless charged KG equation in the charged $C$-metric background can respectively be transformed into theGHEs. We have investigated the asymptotic behaviors of the radial wave functions from both the Heun solutions and the DRS approximations. The two approaches provide the same wave functions asymptotically. We have applied these wave functions to explore Hawking radiation, obtaining the scattering probability and energy density. Furthermore, we have also provided the exact proportionality coefficients in these wave functions.
The wave solutions were then further analysed under different physical boundary conditions, such as those for QNMs and superradiance. We have also obtained the exact wave solutions in terms of local Heun functions and connection coefficients. These enabled us to determine QNMs and superradiance.

Since QNMs are significant for black holes through gravitional waves, we would like to present the numeric results for QNMs based on Sec.\ref{gheeeb} in the future. In \cite{cccmetric,ccccmetric,ccmetric}, the program package QNMspectral developed in \cite{ajo} was applied to do numerical simulations (see also \cite{ebvcas}). 
As per our work from the perspective of Heun equations, we will use the Mathematica package designed to solve Heun equations based on the procedure used in \cite{yh}. It is expected that the analytic expressions of the solutions allow us to make fast numerical calculations of high precision without restrictions on the model parameters, and we would like to show the QNMs' dependence upon the $C$-metric parameters and the charge of scalar fields. 

The methods presented in this paper can be applied to study exact solutions and wave scattering of test fields in other backgrounds. It is also of particular interest to consider Dirac fields in a $C$-metric BH. We hope to examine some of the above problems and report their results elsewhere.

\section*{Acknowledgments}
M.Chen has been supported by the UQ Research Training Program Scholarship. G.Tartaglino-Mazzucchelli has been supported by the Australian Research Council (ARC) Future Fellowship FT180100353, ARC Discovery Project DP240101409, and the Capacity Building Package of the University of Queensland. Y.Z.Zhang has been partially supported by Australian Research Council Discovery Project DP190101529.

\appendix

\section{Derivation of the radial Heun equation}
\label{heun} 
In this appendix, we provide the computation details for deriving the radial Heun 
equation of section \ref{ghe}. 

Setting $\Psi(z)=(z-z_{\infty})^{-1} \Phi(z)$, we transform \eqref{rnc14}  to the form
\begin{eqnarray}\label{rnc15}
 \Phi^{\prime\prime}(z) +\frac{(r_{+}-r_{-})^2z^2_{\infty}}{(z-z_{\infty})^4} \bigg[ \frac{1}{4} \big(\frac{\Delta^{\prime}(r)}{\Delta(r)}\big)^2 - \frac{1}{3} \frac{\Delta^{\prime\prime}(r)}{\Delta(r)}  +\frac{\Omega(r)}{\Delta^2(r)} + \frac{\lambda}{\Delta(r)} \bigg] \Phi(z)= 0. &
\end{eqnarray}
After re-expressing all the functions of $r$  in terms of the new variable $z$ defined in (\ref{rnc8}), we arrive at the following ODE with the singularity  $z=z_{\infty}$ removed,
\begin{eqnarray}\label{rnc16}
\Phi^{\prime\prime}(z) + \bigg\{ \bigg[\frac{A_1}{z}+\frac{A_2}{z-1}+\frac{A_3}{z-z_r}\bigg]   + \bigg[ \frac{B_1}{z^2}+\frac{B_2}{(z-1)^2}+\frac{B_3}{(z-z_r)^2} \bigg] \bigg\} \Phi(z)=0. &
\end{eqnarray}
Here, we have introduced the various constants
\bsubeq
\begin{eqnarray}\label{rnc5051}
A_1&=&
\frac{1}{3}\bigg[ \frac{r^{\p}_{+}-r_{+}}{r^{\p}_{+}-r_{-}} 
+\frac{1}{2}\frac{r_{-}-r_{+}}{r_{-}-r^{\p}_{+}}-\frac{1}{2}\frac{(r_{-}-r_{+})(r^{\p}_{+}-r_{+})}{(r^{\p}_{+}-r_{-})(r^{\p}_{-}-r_{+})} \bigg] 
\nonumber \\
&&+\frac{1}{z^2_r}\bigg(2E+D+\frac{2E}{z_r}\bigg)
-\frac{\lambda}{\alpha^2}\frac{1}{(r_{-} - r^{\p}_{+})(r_{-}-r^{\p}_{-})}\frac{z_{\infty}}{z_r} , 
\end{eqnarray}
\begin{eqnarray}\label{rnc5052}
A_2&=&\frac{1}{3}\bigg[ \frac{r^{\p}_{+}-r_{+}}{r_{+}-r_{-}}-\frac{1}{2}\frac{r_{-}-r^{\p}_{+}}{r_{-}-r_{+}} -\frac{1}{2}\frac{(r_{-}-r^{\p}_{+})(r^{\p}_{+}-r_{+})}{(r^{\p}_{+}-r^{\p}_{-})(r_{-}-r_{+})} \bigg]
\nonumber \\
&& +\frac{1}{(z_r-1)^2}\bigg[2\frac{A+B+C+D+E}{z_r-1} +(2A+B) 
-(2E+D)\bigg]
\nonumber \\
&&
+\frac{\lambda}{\alpha^2}\frac{1}{(r_{-} - r^{\p}_{+})(r_{-}-r^{\p}_{-})}\frac{z_{\infty}}{z_r-1},   
\end{eqnarray}
\begin{eqnarray}\label{rnc5053}
A_3&=&\frac{1}{3}\frac{(r^{\p}_{+}-r_{+})(r^{\p}_{-}-r_{-})}{(r^{\p}_{+}-r_{-})(r_{+}-r_{-})}
+ \frac{1}{6} \frac{(r_{-}-r^{\p}_{-})^2(r^{\p}_{+}-r_{+})}{(r_{-}-r_{+})(r^{\p}_{+}-r_{-})(r^{\p}_{-}-r_{+})}   
    \nonumber \\
&&
+ \frac{1}{6} \frac{(r_{-}-r^{\p}_{-})^2(r^{\p}_{+}-r_{+})}{(r_{-}-r^{\p}_{+})(r^{\p}_{+}-r^{\p}_{-})(r_{-}-r_{+})}
- \frac{1}{z^2_r}\bigg[(2E+D)  
+\frac{2E}{z_r}\bigg]
 \nonumber \\
& &
-\frac{1}{(z_r-1)^2}\bigg[\;2\frac{A+B+C+D+E}{z_r-1} +(2A+B)   
-(2E+D)\bigg]
  \nonumber \\
&&
-\frac{\lambda}{\alpha^2}\frac{1}{(r_{-} - r^{\p}_{+})(r_{-}-r^{\p}_{-})}\frac{z_{\infty}}{z_r}\frac{1}{z_r-1}, 
\end{eqnarray}
\esubeq
with
\bsubeq
\begin{eqnarray}
&&A=\frac{\Omega(r_{-})}{[\Delta^{\p}(r_{-})]^2},\\
&&B=\frac{-4\omega^2r_{-}^3r_{+} + 2qQ\omega (3r_{-}+r_{+}) - 4q^2Q^2}{[\Delta^{\p}(r_{-})]^2}z_{\infty},\\
&&C=\frac{6[\omega^2r_{-}^2r^2_{+} - qQ\omega (r_{-}+r_{+}) + q^2Q^2]}{[\Delta^{\p}(r_{-})]^2}z^2_{\infty},\\
&&D=\frac{-4\omega^2r_{-}r^3_{+} + 2qQ\omega (r_{-}+3r_{+}) - 4q^2Q^2}{[\Delta^{\p}(r_{-})]^2}z^3_{\infty}, \\
&&E=\frac{\Omega(r_{+})}{[\Delta^{\p}(r_{-})]^2} z^4_{\infty}, \\ 
&&\left(\Delta^{\p}(r_{-})\right)^2=\alpha^4 (r_{-}-r_{+})^2(r_{-}-r^{\p}_{+})^2(r_{-}-r^{\p}_{-})^2,
\end{eqnarray}
\esubeq
and 
\bsubeq
\begin{eqnarray}\label{rnc17app3}
B_1-\frac{1}{4}&=&\frac{\Omega(r_{+})}{[\Delta^{\p}(r_{+})]^2},
\\
\label{rnc17app7}
B_2-\frac{1}{4}&=&\frac{\Omega(r^{\p}_{+})}{[\Delta^{\p}(r^{\p}_{+})]^2},
\\
\label{rnc17app8}
B_3-\frac{1}{4}&=& \frac{\Omega(r^{\p}_{-})}{[\Delta^{\p}(r^{\p}_{-})]^2}.
\end{eqnarray}
\esubeq

For convenience, we introduce the quantity $B_4$ which is defined from $A$ above via the relation
\begin{eqnarray}\label{rnc17app9}
B_4-\frac{1}{4}=A=\frac{\Omega(r_{-})}{[\Delta^{\p}(r_{-})]^2}.
\end{eqnarray}
Then, it can be verified that the parameters $A_i$ and $B_i$ satisfy the following relations:
\bsubeq\label{rnc17bs12}
\begin{eqnarray}\label{rnc17app10}
A_1+A_2+A_3=0,
\end{eqnarray}
\begin{eqnarray}\label{rnc17app11}
(A_1+A_3)+(A_1+A_2)z_r=B_1+B_2+B_3-B_4.
\end{eqnarray}
\esubeq

We now make the substitution
\begin{eqnarray}\label{rnc22app}
\Phi(z)=z^{C_1}(z-1)^{C_2}(z-z_r)^{C_3} \mathcal{Y}(z).
\end{eqnarray}
This substitution transforms \eqref{rnc16} into the form
\begin{eqnarray}\label{rnc23}
&&\mathcal{Y}^{\p\p}(z) +\bigg\{  \frac{2C_1}{z}+\frac{2C_2}{z-1}+\frac{2C_3}{z-z_r} \bigg\} \mathcal{Y}^{\p}(z)        
\nonumber \\
&&~~~+\bigg\{ \frac{A_1}{z}+\frac{A_2}{z-1}+\frac{A_3}{z-z_r} + \frac{2C_1C_2}{z(z-1)}  
+\frac{2C_1C_3}{z(z-z_r)}+\frac{2C_2C_3}{(z-1)(z-z_r)}
\nonumber \\
&&
~~~~~~~~+ \frac{B_1+C^2_1-C_1}{z^2} +\frac{B_2+C^2_2-C_2}{(z-1)^2}+\frac{B_3+C^2_3-C_3}{(z-z_r)^2} \bigg\} \mathcal{Y}(z)=0. 
\end{eqnarray}
Choosing $C_j, j=1,2,3$ such that
\bsubeq
\begin{eqnarray}
&& C^2_j-C_j+B_j=0, \label{rnc24}\\
&&\Longrightarrow \;\;C_j=\frac{1}{2} \pm i \sqrt{B_j-\frac{1}{4}}, \label{rnc24app}
\end{eqnarray}
\esubeq
we can eliminate the $1/z^2, 1/(z-1)^2, 1/(z-z_r)^2$ terms in \eqref{rnc23}. 
By means of the relation \eqref{rnc17app10}, we have
\begin{eqnarray}\label{rnc28}
&&\mathcal{Y}^{\p\p}(z)   +   \bigg\{  \frac{2C_1}{z}  +  \frac{2C_2}{z-1}  +   \frac{2C_3}{z-z_r} \bigg\} \mathcal{Y}^{\p}(z)    \nonumber\\
&&~~~+   \bigg\{ \frac{[-(A_1+A_3)  -  (A_1+A_2)z_r+2C_1C_2+2C_1C_3+2C_2C_3]z-q}{z(z-1)(z-z_r)} \bigg\} \mathcal{Y}(z)  =   0, ~~~~~~~
\end{eqnarray}
where $q=2C_1C_3+2C_1C_2 z_r-A_1 z_r$. This equation can be written in the form of the general Heun equation \eqref{rnc36}, i.e.
\begin{eqnarray}\label{rnc36app}
\mathcal{Y}^{\prime\prime}(z)  +   \bigg(   \frac{\gamma}{z} +   \frac{\delta}{z  -  1}  +  \frac{\epsilon}{z  -  z_r}   \bigg) \mathcal{Y}^{\prime}(z)
  +   \frac{\alpha\beta z-q}{z(z  -  1)(z  -  z_r)}  \mathcal{Y}(z)=0,~~~~
\end{eqnarray}
where $\gamma, \delta, \epsilon, q$ are given in \eqref{apprefbs13}, and $\alpha$ and $\beta$ are
\bsubeq
\begin{eqnarray}
\alpha&=&\bigg(C_1+C_2+C_3 -\frac{1}{2}\bigg) 
\nonumber\\
&&
\pm \sqrt{ (C_1^2-C_1)+(C_2^2-C_2)+(C_3^2-C_3)+\frac{1}{4}  +(A_1+A_3)+(A_1+A_2)z_r },~~~~~~
\end{eqnarray}
\begin{eqnarray}\label{rnc35}
\beta&=& \bigg(C_1+C_2+C_3 - \frac{1}{2}\bigg) 
\nonumber\\
&&
\mp \sqrt{ (C_1^2-C_1)+(C_2^2-C_2)+(C_3^2-C_3)+ \frac{1}{4}  + (A_1+A_3)+(A_1+A_2)z_r }.~~~~~~
\end{eqnarray}
\esubeq
It can be easily seen that these parameters satisfy the required condition $\gamma+\delta+\epsilon=\alpha+\beta+1$ for the GHE.

We now show that the $\alpha$ and $\beta$ expressions above are exactly the same as those in \eqref{apprefbs13}. Using the identity \eqref{rnc17app11} and the relation \eqref{rnc24}, the expressions for $\alpha$ and $\beta$ can be simplified to the following
\bsubeq
\begin{eqnarray}\label{rnc300}
\alpha=\bigg(C_1+C_2+C_3 - \frac{1}{2}\bigg)\pm i\sqrt{ B_4-\frac{1}{4}},
\end{eqnarray}
\begin{eqnarray}\label{rnc305}
\beta=\bigg(C_1+C_2+C_3 - \frac{1}{2}\bigg) \mp i\sqrt{B_4-\frac{1}{4}}.
\end{eqnarray}
\esubeq
Introducing $\tilde{B}_j=\sqrt{B_j-\frac{1}{4}}, j=1,2,3,4$, then \eqref{rnc24app} can be written as
\begin{eqnarray}\label{rnc30}
C_j=\frac{1}{2} \pm\tilde{B}_j,\qquad
\tilde{B}_j := i \frac{J(r_{j})}{\Delta^{\p}(r_{j})},
\end{eqnarray}
with the same identification as before: $r_1, r_2, r_3, r_4 \Longleftrightarrow  r_{+},r^{\p}_{+},r^{\p}_{-},r_{-}$. In terms of $\tilde{B}_j$, $\alpha$ and $\beta$ become
\begin{eqnarray}\label{appreff}
\alpha=1\pm \sum_{j=1}^4\tilde{B}_j,  \;\; \beta= 1\pm \sum_{j=1}^3\tilde{B}_j\mp\tilde{B}_4,
\end{eqnarray}
which are nothing but the ones in \eqref{apprefbs13}-\eqref{apppref}. This completes our derivation of \eqref{rnc36}.

The derivation of the angular Heun equation is completely analogue to the derivation above.


\begin{thebibliography}{99}
\bibitem{ligo} B.P.Abbott, et al, ``{\it LIGO Scientific and Virgo Collaborations}'', {\it Phys.\ Rev.\ Lett.}, {\bf 116}: 221101 (2016). \par

\bibitem{raea} R.Abbott, et al, ``{\it GWTC-2:Compact Binary Coalescences Observed by LIGO and Virgo During the First Half of the Third Observing Run}'', {\it Phys.\ Rev.\ X}, {\bf 11}: 021053 (2021). \par

\bibitem{m87} The Event Horizon Telescope Collaboration, ``{\it First M87 Event Horizon Telescope Results. I. The Shadow of the supermassive Black Hole}'', {\it Astrophys.\ J. \ Lett }, {\bf 875}: L1 (2019). \par

\bibitem{hawk4} S.W.Hawking, ``{\it Information loss in black holes}'', {\it Phys.\ Rev.\ D}, {\bf 72}: 084013 (2005). \par

\bibitem{rprm} R.Penrose and R.M.Floyd, ``{\it Extraction of rotational energy from a black hole}'', {\it Nature}, {\bf 229}: 177 (1971). \par

\bibitem{jdb} J.D.Bekenstein, ``{\it Extraction of energy and charge from a black hole}'', {\it Phys.\ Rev.\ D}, {\bf 7}: 949 (1973). \par

\bibitem{kdun} K.Destounis, ``{\it Superradiant instability of charged scalar fields in higher-dimensional Reissner-Nordstr\"{o}m-de Sitter black holes}'', {\it Phys.\ Rev.\ D}, {\bf 100}: 044054 (2019). \par
\bibitem{rbvcpp} R.Brito, V.Cardoso, and P.Pani, ``{\it Superradiance}'', {\it Lect.\ Notes \ Phys.}, {\bf 906}: 1 (2015). \par

\bibitem{ebvccmw} E.Berti, V.cardoso and C.M.Will, ``{\it On gravitational-wave spectroscopy of massive black holes with the space interferometer LISA}'', arXiv: gr-qc/0512160v2 [gr-qc] (2006). \par

\bibitem{ajo} A.Jansen, ``{\it Overdamped modes in Schwarzchild-de Sitter and a Mathematica package for the numerical computation of quasinormal modes}'', {\it Eur. \ Phys.\ J. \ Plus }, {\bf 132}: 546 (2017). \par

\bibitem{kdk} K.D.Kokkotas and B.G.Schmidt, ``{\it Quasinormal modes of stars and black holes}'', {\it Living \ Relativity}, {\bf 2}: 2 (1999). \par

\bibitem{bss} D.Birmingham, I.Sachs, and S.N.Solodukhin, ``{\it Conformal Field Theory Interpretation of Black Hole Quasinormal Modes}'', {\it Phys.\ Rev.\ Lett.}, {\bf 88}: 151301 (2002). \par

\bibitem{ebvcas} E.Berti, V.Cardoso and A.O.Starinets, ``{\it Quasinormal modes of black holes and black branes}'', {\it Class.\ Quantum \ Grav.}, {\bf 26}: 163001 (2009). \par

\bibitem{teu} S.A.Teukolsky, ``{\it Perturbations of a rotating black hole. I. Fundamental equations for gravitational, electromagnetic, and neutrino-field perturbations}'', {\it Astrophys.\ J.}, {\bf 185}: 635-647 (1973). \par

\bibitem{teu2} W.H.Press and S.A.Teukolsky, ``{\it Perturbations of a rotating black hole. II. Dynamical stability of the Kerr metric}'', {\it Astrophys.\ J.}, {\bf 185}: 649-673 (1973). \par

\bibitem{dvgn} D.V.Gal'tsov and D.N\'{u}$\tilde{\textrm{n}}$ez, ``{\it Exact solutions to the First-order perturbation problem in a de Sitter background}'', {\it Gen.\ Rel. \ Grav.}, {\bf 21}: 3 (1989). \par

\bibitem{yhhz} Y.Huang and H.S.Zhang, ``{\it Quasibound states of charged dilatonic black holes}'', {\it Phys.\ Rev.\ D}, {\bf 103}: 044062 (2021). \par

\bibitem{hskd} H.S.Vieira and K.D.Kokkotas, ``{\it Quasibound states of Schwarzchild acoustic black holes}'', {\it Phys.\ Rev.\ D}, {\bf 104}: 024035 (2021). \par

\bibitem{hsvc2} H.S.Vieira, V.B.Bezerra and C.R.Muniz, ``{\it Exact solutions of the Klein-Gordon equation in the Kerr-Newman background and Hawking radiation}'', {\it Ann.\ Phys.}, {\bf 350}: 14-28 (2014). \par

\bibitem{hsvc4} H.S.Vieira and V.B.Bezerra, ``{\it Charged scalar fields in a Kerr-Sen black hole: exact solutions, Hawking radiation, and resonant frequencies}'', {\it Chin. \ Phys.\ C }, {\bf 43}: 035102 (2019). \par

\bibitem{sqw} S.Q.Wu and X.Cai, ``{\it Massive complex scalar field in the Kerr-Sen geometry: Exact solution of wave equation and Hawking radiation} '', {\it J. \ Math.\ Phys. }, {\bf 44}: 3 (2003). \par

\bibitem{masadd} M.Nozawa and T.Kobayashi, ``{\it Quasinormal modes of black holes localized on the Randall-Sundrum 2-brane} '', {\it arXiv:}\href{https://arxiv.org/pdf/0803.3317.pdf}{ 0803.3317 (2008)}. \par

\bibitem{heun} A.Ronveaux, ``{\it Heun's Differential Equations}'', Oxford University, New York, (1995). \par

\bibitem{smhset} S.Mano, H.Suzuki and E.Takasugi, ``{\it Analytic solutions of the Teukolsky equation and their low frequency expansions}'', {\it Prog.\ Theor. \ Phys.}, {\bf 95}: 6 (1996). \par

\bibitem{hseh} H.Suzuki, E.Takasugi and H.Umetsu, ``{\it Perturbations of Kerr-de Sitter black hole and Heun's equations}'', {\it Prog.\ Theor.\ Phys.}, {\bf 100}: 3 (1998). \par

\bibitem{dbhs} D.Batic and H.Schmid, ``{\it Heun equation, Teukolsky equation, and Type-D matrics}'', {\it J. \ Math.\ Phys. }, {\bf 48}: 042502 (2007). \par

\bibitem{mhort} M.Hortacsu, ``{\it The radial Teukolsky equation for Kerr-Newman-de Sitter geometry: Revisited}'', {\it Eur. \ Phys.\ J. \ Plus }, {\bf 136}: 13 (2021). \par

\bibitem{bgadd2} J.B.Amado and B.Gwak, ``{\it Scalar quasi-normal modes of accelerating Kerr-Newman-AdS black holes}'',  {\it arXiv:}\href{https://arxiv.org/abs/2309.11355}{ 2309.11355 (2023)}. \par

\bibitem{hsvc3} H.S.Vieira and V.B.Bezerra, ``{\it Confluent Heun functions and the physics of black holes: resonant frequencies, Hawking radiation and scattering of scalar waves}'', {\it Ann.\ Phys.}, {\bf 373}: 28-42 (2016). \par

\bibitem{hmsn} H.Motohashi and S.Noda, ``{\it Exact solution for wave scattering from black holes: Formulation}'', {\it Prog.\ Theor.\ Exp. \ Phys.}, \href{https://inspirehep.net/files/48f294c43cf60fca490cfb3bc4719381}{{\bf 083E03} (2021)}. \par

\bibitem{snhm} S.Noda and H.Motohashi, ``{\it Spectroscopy of Kerr-$AdS_5$ spacetime with the Heun function: Quasinormal modes, greybody factor, and evaporation}'', {\it Phys.\ Rev.\ D}, {\bf 106}: 064025 (2022). \par

\bibitem{yh} Y.Hatsuda, ``{\it Quasinormal modes of Kerr-de Sitter black holes via the Heun function}'', {\it Class.\ Quantum \ Grav.}, {\bf 38}: 025015 (2021). \par

\bibitem{ccmetric} K.Destounis, G.Mascher and K.D.Kokkotas, ``{\it Dynamical behavior of the C-metric: Charged scalar fields, quasinormal modes, and superradiance}'', {\it Phys.\ Rev.\ D}, \href{https://journals.aps.org/prd/abstract/10.1103/PhysRevD.105.124058}{{\bf 105}: 124058 (2022)}. \par

\bibitem{jb} J.B.Griffiths, P.Krtou\v{s} and J.Podolsk\'{y}, ``{\it Interpreting the C-metric}'', {\it Class.\ Quantum \ Grav.}, {\bf 23}: 6745-6766 (2006). \par

\bibitem{kh} K.Hong and E.Teo, ``{\it A new form of the C-metric}'', {\it Class.\ Quantum \ Grav.}, {\bf 20}: 3269-3277 (2003). \par

\bibitem{Klemm:2013eca}
D.~Klemm and M.~Nozawa,
``{\it Supersymmetry of the C-metric and the general Plebanski-Demianski solution}'',
{\it J.\ High \ Ener. \ Phys.}, {\bf 05}: 123 (2013). \par

\bibitem{Nozawa:2022upa} M.~Nozawa and T.~Torii, ``{\it New family of C metrics in N=2 gauged supergravity}'', {\it Phys.\ Rev.\ D}, {\bf 107}: 064064 (2023). \par

\bibitem{Nozawa:2023aep} M.~Nozawa and T.~Torii, ``{\it Wormhole C metric}'', {\it Phys.\ Rev.\ D}, {\bf 108}: 064036 (2023). \par

\bibitem{cccmetric} K.Destounis, R.D.B. Fontana and F.C.Mena, ``{\it Accelerating black holes: Quasinormal modes and late-time tails}'', {\it Phys.\ Rev.\ D}, {\bf 102}: 044005 (2020). \par

\bibitem{ccccmetric} K.Destounis, R.D.B. Fontana and F.C.Mena, ``{\it Stability of the Cauchy horizon in accelerating black-hole spacetimes}'', {\it Phys.\ Rev.\ D}, {\bf 102}: 104037 (2020). \par

\bibitem{bgadd} B.Gwak, ``{\it Quasinormal modes in near-extremal spinning $C$-metric}'', {\it Eur. \ Phys.\ J. \ P.}, \href{https://link.springer.com/content/pdf/10.1140/epjp/s13360-023-04215-7}{{\bf 138}: 582 (2023)}. \par

\bibitem{dkfn} D.Kofron, ``{\it Separability of test fields equations on the C-metric background}'', {\it Phys.\ Rev.\ D}, {\bf 92}: 124064 (2015). \par

\bibitem{synutf} S.Yoshida, N.Uchikata and T.Futamase, ``{\it Quasinormal modes of Kerr-de Sitter black holes}'', {\it Phys.\ Rev.\ D}, {\bf 81}: 044005 (2010). \par

\bibitem{azhi} A.Zhidenko, ``{\it Quasi-normal modes of Schwarzchild-de Sitter black holes}'', {\it Class.\ Quant.\ Grav.}, {\bf 21}: 273-280 (2004). \par

\bibitem{bbacc} D.Bini, C.Cherubini and A.Geralico, ``{\it Massless field perturbations of the spinning C-metric}'', {\it J. \ Math.\ Phys. }, {\bf 49}: 062502 (2008). \par

\bibitem{viqbs} H.S.Vieira, K.Destounis and K.D.Kokkotas, ``{\it Slowly-rotating curved acoustic black holes: Quasinormal modes, Hawking-Unruh radiation and quasibound states}'', arXiv: 2112.08711v2 [gr-qc] (2022). \par

\bibitem{dr} T.Damour and R.Ruffini, ``{\it Black-hole evaporation in Klein-Sauter-Heisenberg-Euler formalism}'', {\it Phys.\ Rev.\ D}, {\bf 14}: 2 (1976). \par

\bibitem{page2} D.N.Page, ``{\it Particle emission rates from a black hole. II. Massless particles from a rotating hole}'', {\it Phys.\ Rev.\ D}, {\bf 14}: 3260 (1976). \par

\bibitem{s} S.Sannan, ``{\it Heuristic Derivation of the Probability Distributions of Particles Emitted by a Black Hole}'', {\it Gen.\ Relat. \ Grav.}, {\bf 20}: 3 (1988). \par

\bibitem{miladd} A.Ball and N.Miller, ``{\it Accelerating Black Hole Thermodynamics with Boost Time} '', {\it arXiv:}\href{https://arxiv.org/abs/2008.03682}{ 2008.03682 (2021)}. \par

\bibitem{miladd2} A.Ball, ``{\it Global First Laws of Accelerating Black Holes} '', {\it arXiv:}\href{https://arxiv.org/abs/2103.07521}{ 2103.07521 (2021)}. \par

\bibitem{edwl} E.W.Leaver, ``{\it Quasinormal modes of Reissner-Nordstr\"{o}m black holes}'', {\it Phys.\ Rev.\ D}, {\bf 41}: 10 (1990). \par

\bibitem{nacjh} N.Adersson and C.J.Howls, ``{\it The asymptotoc quasinormal mode spectrum of non-rotating black holes}'', {\it Class.\ Quant.\ Grav.}, {\bf 21}: 1623-1642 (2004). \par

\bibitem{ebkdk} E.Berti and K.D.Kokkotas, ``{\it Quasinormal modes of Reissner-Nordstr\"{o}m-anti-de Sitter black holes: Scalar, electromagnetic, and gravitational perturbations}'', {\it Phys.\ Rev.\ D}, {\bf 67}: 064020 (2003). \par

\bibitem{carroll} S.M.Carroll, ``{\it Spacetime and Geometry}'',  {\it Integre Technical Publishing CO., USA}, (2004). \par

\bibitem{erpb} E.D.Rainville and P.E.Bedient, ``{\it Elementary Differential Equations}'', Macmillan Publishing CO., Inc., New York, (1974). \par

\bibitem{mgigm} M.Giammatteo and I.G.Moss, ``{\it Gravitational quasinormal modes for Kerr anti-de Sitter black holes}'', {\it Class.\ Quant.\ Grav.}, {\bf 22}: 1803-1824 (2005). \par

\bibitem{fwdpcl} F.Willenborg, D.Philipp and C.L\"{a}mmerzahl, ``{\it Exact wave-optical inaging of a Kerr-de Sitter black hole using Heun's equation}'', arXiv: 2310.12917v1 [gr-qc] (2023). \par









\end{thebibliography}
\end{document}